\documentclass[prd,showpacs,onecolumn,preprintnumbers
]{revtex4}

\usepackage{amsmath} \usepackage{graphicx} \usepackage{amsfonts}
\usepackage{array} \usepackage{amsthm} \usepackage[cp1251]{inputenc}
\usepackage[russian]{babel}
\usepackage{psfrag}

\usepackage{amsmath}
\newcommand{\btt}{\tilde{\beta}}

\newcommand{\ben}{\begin{enumerate}}
\newcommand{\een}{\end{enumerate}}

\newcommand{\la}{\lambda}

\newcommand{\al}{\alpha}

\newcommand{\bw}{\begin{widetext}}
\newcommand{\ew}{\end{widetext}}

\newcommand{\be}{\begin{equation}}
\newcommand{\ee}{\end{equation}}
\newcommand{\ba}{\begin{eqnarray}}
\newcommand{\ea}{\end{eqnarray}}
\newcommand{\non}{\nonumber}

\newcommand{\eps}{\varepsilon}

\def\cF{{\cal F}}

\def\cG{{\cal G}}
\def\cR{{\cal R}}

\def\e{{\rm e}}
\newcommand{\ec}{\ensuremath{\varepsilon_{c}}}
\newcommand{\ve}{\varepsilon}
\newcommand{\bei}{\begin{itemize}}
\newcommand{\eei}{\end{itemize}}

 \newcommand{\nn}{\nonumber}
\newcommand{\qd}{\qquad} \newcommand{\ra}{\rightarrow}
\newcommand{\Lef}{\left(} \newcommand{\Rig}{\right)}

 \newcommand{\ep}{\varepsilon}
 
\newcommand{\Vi}{V_{\rm int}} 
\begin{document}
\title{Cosmological models with gauge fields}

\author{Dmitry V. Gal'tsov}\email{galtsov@physics.msu.ru}
\affiliation{Department of Theoretical Physics, Moscow State
University, 119899, Moscow, Russia}
\author{Evgeny A. Davydov} \email{eugene00@mail.ru} \affiliation{Bogoliubov
Laboratory of Theoretical Physics, JINR, 141980, Dubna, Moscow
region, Russia}
\begin{abstract}
We discuss cosmological models  involving homogeneous and isotropic
Yang-Mills (YM) fields. Such models  were proposed recently as an
alternative to scalar models of cosmic acceleration. There exists a
unique $SU(2)$ YM configuration (generalizable to larger gauge
groups) whose energy-momentum tensor is homogeneous and isotropic in
space. It is parameterized by a single scalar field with a quatric
potential. In the case of the closed universe the coupled YM --
doublet Higgs system  admits homogeneous and isotropic
configurations too. While pure Einstein-Yang-Mills (EYM) cosmology
with the standard conformally invariant YM action gives rise to the
hot universe, Einstein-Yang-Mills-Higgs (EYMH) cosmology has a
variety of regimes which include inflationary stages, bounces, and
exhibits global cycling behavior reminiscent of the Multiverse
developed in time. We also discuss other mechanisms of conformal
symmetry breaking such as string-inspired Born-Infeld (BI)
modification of the YM action or field-theoretical quantum
corrections.
\end{abstract}
\maketitle
\section{Introduction}
With  discovery of inflation as  solution of the horizon and
flatness problems in cosmology \cite{Gu} it became widely accepted
that, apart from gravity, some other homogeneous and isotropic
fields have to be present at  cosmological scale which mimic
variable cosmological constant. Traditionally this role is
attributed to a single scalar field, the inflaton, or several scalar
fields \cite{rews}. Modern theories provide various candidates for
inflaton varying from  Higgs field of the standard model to more
hypothetical moduli fields originating from compactified
supergravity/string theory. Still, physical nature of the inflaton
is far from being uniquely understood, and the choice of the
inflaton potential remains mostly phenomenological. Similarly,
popular current models of dark energy \cite{DE} involve scalar
fields with rather exotic properties (quintessence,
$\kappa$-essence, phantom) whose physical nature is far from clear.
Moreover, no massless elementary scalar field was observed
experimentally so far. Meanwhile, vector fields certainly do exist
(and are masseless before the spontaneous symmetry breaking) being
basic ingredients of the Standard Model and its generalization.
Therefore an idea to use vector fields instead or together with
scalar ones to model inflation and dark energy seems to be
appropriate. In fact, the model of inflation driven by vector field
was suggested long ago \cite{Ford}, but remained unnoticed by
cosmologists until recently when it was revived in the context of
the dark energy problem \cite{Arm}. Formation of YM condensates in
superdense matter was discussed long ago by Linde
\cite{Linde:1979pr}.

There are two major reasons why vector fields were not welcome in
cosmology, apart from their relative complexity. Spatially
homogeneous configuration of a single (Abelian) vector field
evidently can not be isotropic. Therefore, in order to fit the
Friedmann-Robertson-Walker (FRW) cosmology one has to introduce (at
least) a triplet of vector fields ensuring isotopy of the total
energy-momentum tensor. Another problem is conformal invariance of
the standard classical YM lagrangian, which leads within the FRW
cosmology to the equation of state $w=P/\ep=1/3$, identical with
that of the photon gas. Therefore, the solution for the coupled EYM
system will be the hot Universe driven by the cold classical matter
field \cite{Galtsov:1991un}. Meanwhile, for an accelerated expansion
one needs the equation of state $-1\le w\le -1/3$, so the conformal
invariance must be broken.

However, the first problem is avoided in  the non-Abelian case: the
SU(2) triplet of YM fields admits an essentially non-Abelian
configuration (with non-zero commutator of the matrix-valued
potentials) whose stress-tensor exhibits three-dimensional
homogeneity and isotropy. The second problem (conformal symmetry
breaking) can be overcome passing to effective actions which account
for quantum corrections either in the context of gauge theories or
string theory. Various attempts to use YM vector fields in
constructing dark energy models thus were undertaken during recent
years \cite{YMDE}.

\section{SU(2)-driven FRW cosmology}
 Consider the FRW interval in the conformal gauge
 \be
 ds^2=a^2(\eta)\left(d\eta^2-dl_k^2\right)
 \ee
 where $k=-1,\,0,\,1$
for open, flat and closed spatial geometry.
 As it was shown for   $k=1,0$ by  Cervero and Jacobs
\cite{Cervero:1978db},   Henneaux \cite{Henneaux:1982vs} and
Hosotani \cite{Hosotani:1984wj} and  for all   $k$ by   Gal'tsov and
Volkov \cite{Galtsov:1991un}, the following configuration
 \be
E^a_i=\dot f \;\delta^a_i,\quad B^a_i= (k-f^2)\; \delta^a_i, \ee
parametrized by the single scalar function $f(\eta)$ of the
conformal time gives rise to homogeneous and isotropic
energy-momentum tensor. The YM lagrangian density in the conformal
frame then reads
  \be
L=\frac{3}{8\pi a^4}\left({\dot f}^2-(k-f^2)^2 \right)
 \ee
 so the electric part corresponds to
the kinetic term, while  the magnetic part --- to the potential term
in the action. The effective scalar field $f$ is dimensionless (we
use the units $\hbar=c=1$), so in spite of similarity with the
scalar field theory with the quatric potential,  dependence of the
lagrangian density on the  scale factor $a$ is different: the energy
scales as $1/a^4$ which is characteristic for the conformal field.
The energy-momentum tensor is traceless and the equation of state is
 \be P=\ep/3\quad {\rm with}\quad
\ep=\frac3{8\pi a^4} \left[{\dot f}^2+(k-f^2)^2\right].
 \ee
Thus one obtains the hot Universe driven by non-thermal matter
\cite{Galtsov:1991un}: our classical YM configuration  perfectly
mimics the photon gas.

The YM equations reduce to equations of motion of a fictitious
particle in the potential well
 \be W_k= (k-f^2)^2\vspace{-.2cm}
 \ee
which   is especially interesting in the closed case $k=1$ when it
is the double well potential (Fig.1). Two minima correspond to
neighboring topologically different vacua. We therefore observe that
gravity lowers the potential barrier between topological sectors to
a finite value, similarly to the Higgs field. Physically this
similarity is due to attractive nature of both (contrary to
repulsive nature of YM). If the energy is less than the height of
the potential barrier, the particle oscillates  around a single
vacuum, when it is above the barrier, oscillations between different
vacua are unsuppressed.

In the flat and open cases the potential has one minimum at $f=0$
with $W=0$ in the flat case $k=0$ and $W=1$ in the open case $k=-1$.
so there are no topological effects.

Computing the Chern-Simons 3-form
 \be
 \omega_3=\frac{e^2}{8\pi^2} {\rm Tr} \left(
A\wedge dA -\frac{2ie}3 A\wedge A\wedge A\right),
 \ee satisfying the equation
  \be
d\omega_3= \frac{e^2}{8\pi^2} {\rm Tr} F\wedge F,
 \ee
one finds that it is non-trivial in the closed case $k=1$, giving
the winding number of the map $SU(2)\to S^3$:
 \be
  N_{CS}=\int_{S^3} \omega_3=\frac14
(f+1)^2(2-f).
 \ee
 The vacuum $f=-1$ is topologically trivial: $N_{CS}=0$,
while the vacuum $f=1$ is the next non-trivial one with $N_{CS}=1$.

Generalization of the above ansatz  for $SU(n)$ and $SO(n)$ gauge
groups and further classical and quantum properties of EYM
cosmological solutions were considered in a number of papers
\cite{Moni,Ku} in the  90-ies. Behavior of small perturbations in
cosmologies with vector fields was discussed more recently in
\cite{Inst}.
\begin{figure}
\center \psfrag{2}{\raisebox{-0.1cm}{\hspace{-0.1cm}\footnotesize
-1}} \psfrag{1}{\raisebox{-0.1cm}{\hspace{-0.1cm}\footnotesize 1}}
\psfrag{inst}{\raisebox{-0cm}{\hspace{-0.5cm}\footnotesize
INSTANTON}} \psfrag{WH}{\raisebox{0cm}{\hspace{0.1cm}\footnotesize
WH}} \psfrag{C}{\raisebox{-0.1cm}{\hspace{-0.3cm}\small C}}
\psfrag{w}{\raisebox{0cm}{\hspace{-0.2cm}\normalsize $f$}}
\psfrag{Vw}{\raisebox{0cm}{\hspace{-0.1cm}\normalsize $W_1(f)$}}
\psfrag{CFRW}{\raisebox{6cm}{\hspace{-1.2cm}\normalsize Closed FRW
Universe}}
%\psfrag{CFRW}{\raisebox{0cm}{\hspace{-1.2cm}\normalsize Closed FRW Universe}}
%\raisebox{-15cm}
{\includegraphics[width=7cm]{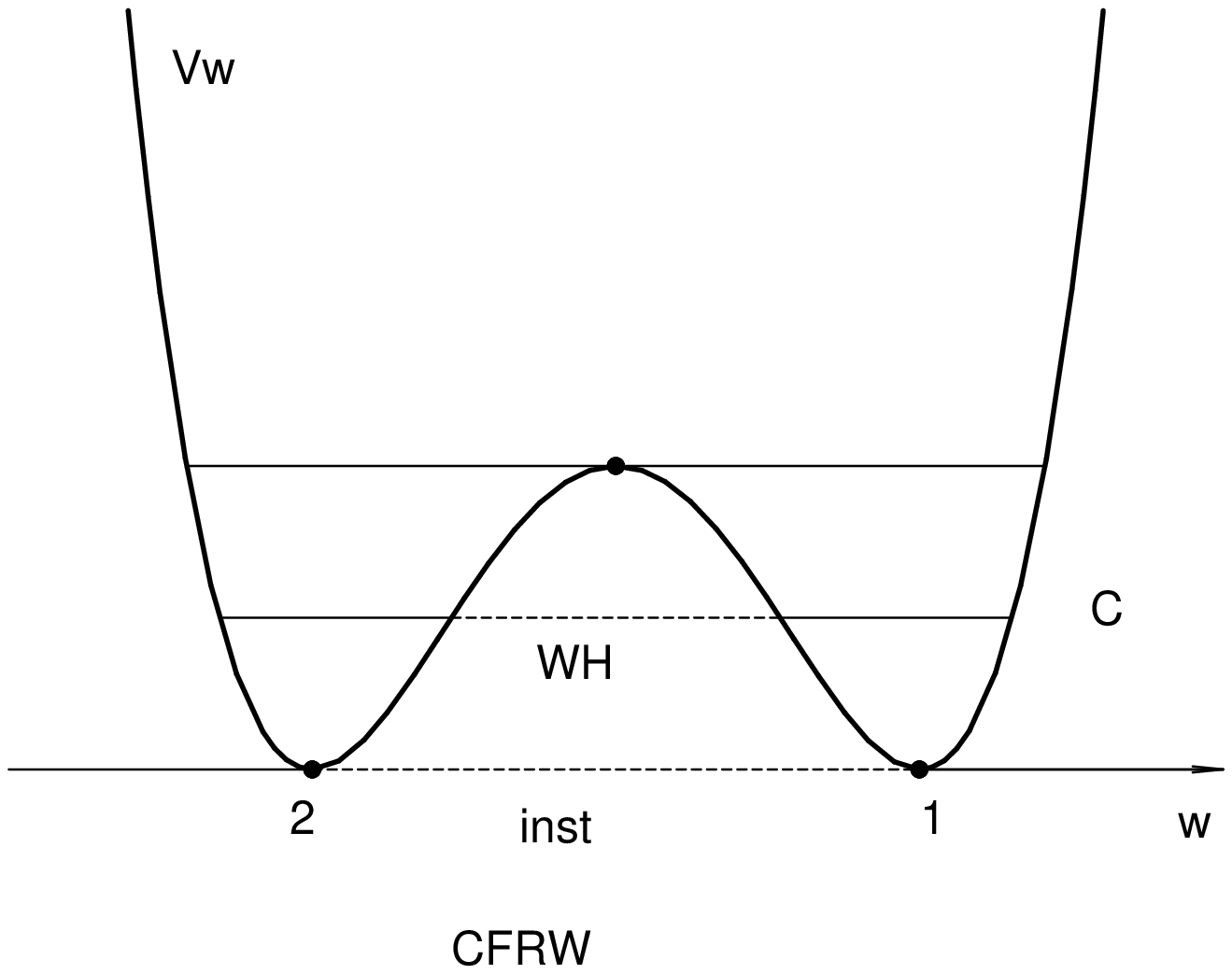}}
%\caption{YM potential, closed universe}
\end{figure}
\subsection{Cosmological sphaleron}
In the closed case there exists a particularly simple configuration
$f=0$ ($f$-particle sitting at the top of the barrier between two
vacua) which in analogy with the sphaleron  in the Weinberg-Salam
(WS) theory was called the ``cosmological sphaleron''~\cite{Gist}.
It is worth noting that the localized particle-like EYM solutions
similar to the WS sphalerons exist as well  which are asymptotically
flat regular particle-like solutions of the EYM equations discovered
by Bartnik and McKinnon (for a review and further references see
\cite{VoGa}). The repulsive YM stresses in these objects are
compensated by gravity instead of the Higgs field in the WS
sphalerons. Creation and decay of sphalerons generates a transition
of the YM field between topological sectors, and it is accompanied
by the fermion number non-conservation in presence of fermions
\cite{VoGa}. The cosmological sphaleron has the topological charge
$N_{CS}=1/2$ like the sphaleron in the Weinberg-Salam theory.

The equation of motion of the $f$-particle \be  \ddot f=2f(1-f^2)
\ee is solved indeed by $f=0$, this correspond to the  total energy
\be {\dot f}^2+(1-f^2)^2=1.\ee The YM field in this case is purely
magnetic. A more general solution with the same energy describes
rolling down of the $f$-particle (the sphaleron decay)~\cite{Gist}:
\be f=\frac{\sqrt{2}}{\cosh(\sqrt{2}\eta)}.\ee  Rolling down to the
vacuum $w=1$ takes an infinite time, while the corresponding full
cosmological evolution is given by \be a=\sqrt{\frac{4\pi G}{g^2}}
\sin\eta \ee and takes a finite time. Thus the cosmological
sphaleron is quasi-stable. This conclusion is not modified if a
positive cosmological constant is added.
\subsection{Instantons and wormholes}
An homogeneous and isotropic EYM system has interesting features
also in the $k=1$ space of Euclidean signature which is invoked in
the path-integral formulation of quantum gravity. Actually, when
$|f|<1$, transitions between two topological sectors can be effected
via underbarrier tunneling described by instanton and wormhole
Euclidean solutions. In the Euclidean regime   the first integral of
the equations of motion reads: \be\dot f^2-(f^2-1)^2=-C, \ee where
$C$ is an integration constant. Instanton corresponds to $C=0$, it
describes tunneling between the vacua $f=\pm 1$. It is a self-dual
Euclidean YM configuration for which the stress tensor is zero,
therefore a conformally flat gravitational field can be added just
as a background.

Tunneling solutions at higher excitation levels $0<C\leq 1$ are not
self-dual. In flat space-time they are known as a meron ($C=1$, the
Euclidean counterpart of the cosmological sphaleron, $N_{CS}=1/2$)
and nested merons $0<C< 1, \;1/2 <N_{CS}<1$.  The energy-momentum
tensor of the meron is non-zero, and in flat space this solution is
singular. When gravity is added, the singularity at the location of
a meron expands to a wormhole throat, and consequently, the
Euclidean topology of the space-time transforms to that of a
wormhole. Topological charge of the meron wormholes is zero, the
charge of the meron being swallowed by the wormhole~\cite{worm}.
 The total action of these wormholes diverges because of slow
fall-off of the meron field at infinity, so the amplitude of
creation of  baby universes associated with the Euclidean wormholes
is zero. However, when a positive cosmological constant is added
(inflation) the action becomes finite due to compactness of the
space. Such solutions can be interpreted as describing tunneling
between the de Sitter space and the hot FRW universe.

  Adding the positive cosmological constant, we will get similar
first integrals both for the $f$-particle and the cosmological
radius: \be\dot f^2-(f^2-1)^2=-C,\quad \dot a^2+(\Lambda
a^4/3-a^2)=-C/(e^2 m_{Pl}^2).\ee Solutions describe independent
tunneling of   $f$  and   $a$ with different periods $T_f,\; T_a$
depending on the excitation level $C$. To be wormholes, they must
obey a quantization condition $n_f T_f=n_a T_a$~\cite{Verb} with two
integers. However, for a specific value of the cosmological constant
\be\Lambda=\frac34m_{Pl}^2, \ee it was found~\cite{Don} that
$T_a=T_f$ for all $C\in [0,1]$. In particular, for $C=1$ (the meron
limit) the radius $a$ becomes constant (Euclidean static Einstein
Universe). For $C\neq 1,0$ the solutions describe creation of the
baby universes (which was invoked in the Coleman's idea of the ``Big
fix''). Remarkably, under the above conditions, the total action
(gravitational plus YM) is precisely zero~\cite{Don}:\be
S_{YM}+S_{gr}=0. \ee Thus, the pinching off of baby universes occurs
with unit probability. For later work on EYM wormholes see
\cite{worm2}.

\section{Einstein-Yang-Mills-Higgs cosmology}
Consider now the EYMH action with complex doublet Higgs:
 \be
    S=
     \int \left\{-\frac{1}{16\pi G}R
      -\frac14  F^a_{\mu\nu}F^{a\mu\nu} +
    \frac{1}{2}(D_\mu \Phi)^\dag D_\mu \Phi-\frac{\lambda}{4}\left(\Phi^\dag\Phi-v^2\right)^2\right\}\sqrt{-g}d^4
    x,
 \ee
where \be \label{eq:cov1}
    D_\mu\Phi=\partial_\mu\Phi+eA^a_\mu T_a\Phi.
\ee It is easy to see that in the case of spatially closed FRW
cosmology an ansatz for Higgs
 \be \label{hig}
      \Phi=h(t)\e^{i\xi(t)}U\Phi_0,
      \quad U=\e^{2\chi T_a n^a},\quad \Phi_0^\dag\Phi_0=1,
  \ee
 is compatible with homogeneity and isotropy of the full EYMH
 system. Indeed, with the YM parametrization
 \be \label{amu}
      e A_\mu^a T_a=\frac{1-f(t)}2\,U\,\partial_\mu U^{-1},
 \ee
  the covariant derivative reduces to the ordinary one
 \be
    D_\mu\Phi=\frac{1+f}2\partial_\mu\Phi.
\ee  Thus an  homogeneous and isotropic  cosmology does exist for
Higgs in the fundamental representation. On the contrary, the
triplet Higgs leads to anisotropic cosmology \cite{bfm}.

The Higgs phase rotation factor $\e^{i\xi(t)}$, which is compatible
with the desired symmetry, in the flat space leads to infinite
energy and must be omitted, but for the closed FRW cosmology its
contribution is finite, so we may keep it.

  The EYMH action contains three different mass parameters: the
Planck mass, the mass of the W-boson, and the Higgs mass
 \be
 M_{\rm Pl}=\frac1{\sqrt{G}},\qquad M_W=ev,\qquad M_H=\sqrt{\la} v.
 \ee
We then rescale the Higgs function $h\to h M_{Pl}$ and introduce the
dimensionless parameters
 \be
\alpha=\frac{M_W}{eM_{\rm Pl}}=\frac{v}{M_{\rm Pl}},\qquad
\beta=\frac{M_H}{M_W}=\frac{\sqrt{\la}}{e}.
 \ee
Finally, keeping in mind that the pure EYM system has a natural
lengths scale $l=1/(e M_{\rm Pl})$, we present the metric in terms
of the dimensionless lapse and scale functions $N,\, a$:
 \be
 ds^2=l^2\left\{-N^2 dt^2+a^2\left[d\chi^2+\sin\chi^2
 (d\theta^2+\sin^2\theta d\varphi^2)\right]\right\}.
  \ee
Then substituting the ansatze (\ref{hig},\ref{amu}) and integrating
over the three-sphere we obtain the one-dimensional reduced action
 \ba
S_1&=&\int\left[\frac3{8\pi}\left(aN-\frac{\dot{a}^2a}{N}\right)+
\frac{3\dot{f}^2 a}{2N}+\frac{\dot{h}^2
a^3}{2N}+\frac{h^2\dot{\xi}^2 a^3}{2N}-V_f-V_h-V_{\rm int} \right]dt\\
V_f&=&\frac{3N(f^2-1)^2}{2a},\quad
V_h=\frac{\beta^2}4(h^2-\alpha^2)^2Na^3,\quad V_{\rm int}=\frac34
h^2(f+1)^2 Na
 \ea
where we have omitted an overall factor $\pi/4$ and the total
derivative in the scalar curvature term.

The equation for the Higgs phase rotation variable $\xi$
\begin{equation}\label{eq:extraLag}
    \frac{d}{dt}\left(\frac{h^2 a^3\dot{\xi}}{N}\right)=0
\end{equation}
can be integrated
\begin{equation}\label{eq:xi}
    \dot{\xi}=\frac{\sqrt{2}jN}{h^2 a^3},
\end{equation}
where $j$ is the integration constant. Then it is easy to check that
the remaining dynamics can be derived from the action in which the
$\xi$-kinetic term is replaced by the potential term
\begin{equation}
    V_j=\frac{j^2}{Nf^2a^3}.
\end{equation}
 Variation with respect to $N$  leads to the constraint
 equation
\begin{equation}\label{eq:N}
    a(\dot{a}^2+1)=\frac{8\pi }{3}\eps,
\end{equation}
and it is convenient to fix the gauge $N=1$ afterwards. For the
energy density we then get
\begin{equation}\label{eq:rho}
    \eps=T_f+T_h+V_f+V_h+\Vi+V_j,
\end{equation}
where the first two terms are kinetic:
 \be T_h=\frac12\dot{h}^2 a^3, \qquad T_f=\frac32\dot{f}^2,
 \ee
and the remaining are four potentials:
 \be
 V_f=\frac{3(f^2-1)^2}{2a},\quad
V_h=\frac{\beta^2}4(h^2-\alpha^2)^2a^3,\quad V_{\rm int}=\frac34
h^2(f+1)^2 a,\quad V_j=\frac{j^2}{f^2a^3},
 \ee
The acceleration equation is obtained by variation of the action
with respect to $a$ with account for the constraint:
\begin{equation}\label{eq:a2}
    a^2\ddot{a}=-\frac{4\pi }{3}(\eps+3p),
\end{equation}
leading to the following expression for pressure:
\begin{equation}\label{eq:p}
    p=T_h-V_h+V_j+\frac{T_f}{3}+\frac{V_f}{3}-\frac{\Vi}{3},
\end{equation}
and therefore,
\begin{equation}\label{eq:a3}
    a^2\ddot{a}=-\frac{8\pi }{3}(2T_h+2V_j+T_f-V_h+V_f).
\end{equation}
The field equations for the YM and Higgs scalar functions read
\begin{eqnarray}
  \ddot{f} + 3\frac{\dot{a}}{a}\dot{f} &=& -\frac{3}{2a^2}h(f+1)^2-
  \beta^2(h^2-\al^2)h+\frac{2j^2}{h^3a^6}, \nn\\
  \ddot{f}+\frac{\dot{a}}{a}\dot{f} &=&
  -\frac{3}{2}h^2(f+1)-\frac{6}{ a^2}(f^2-1)f.\label{eq:w}
\end{eqnarray}

From the acceleration equation is clear that kinetic terms always
produce deceleration, while the potential terms are partly
accelerating. In the transient regimes when different potential
terms dominate, one has the following equations of state:

\extrarowheight=3pt
\begin{center}
\begin{tabular}{|l|l|l|}
  \hline
  Dominant potential & $w=p/\eps$ & type \\
  \hline
  Higgs $V_h$ & $w=-1$ & cosmological constant \\
  Higgs phase rotation $V_j$ & $w=1$ & stiff matter \\
  YM potential $V_f$ & $w=1/3$ & radiation\\
  Interaction term $\Vi$ & $w=-1/3$ & string gas \\
  \hline
\end{tabular}
\end{center}
\extrarowheight=0pt

\subsection{Standard model scale}
Behavior of solutions essentially depends on the  parameters
$\alpha,\,\beta$. Consider first the scales relevant to the Standard
model. In this case   $\alpha\sim 10^{-17}$ and $\beta$ is of the
order of unity. The corresponding dynamical scale if far from the
Planck scale and matter contribution to the action is of the order
of   $\alpha^4\beta^2$. The corresponding values of the scale factor
must be of the order $1/\alpha^2\beta$. Then the main contribution
to evolution of the scale factor will come from the Higgs potential
$V_h$. Typical regimes of the evolution of the Higgs scalar
correspond to motion near the extremal points of the potential: the
minima $|h|=\alpha$ and the local maximum $h=0$. In the first case
one observes small oscillations, in the second -- slow rolling down
from the top of the potential barrier. While the frequency of
oscillations is proportional to  $\beta$, the rolling down velocity
is much less, namely of the order of $\alpha^2\beta$.  A substantial
variation of the scale factor will correspond to time intervals of
the order of  $1/\alpha^2\beta$, so the cosmic acceleration will be
proportional to  the average in time values of potential and kinetic
energy of the scalar field $<V_h-2T_h>$. For harmonic oscillations
$<T_h>=<V_h>$, so the acceleration will be negative. The rolling
down regime is exponential, so the characteristic time $T_{roll}$
will depend on initial conditions  as $-\ln h(0)$. During the
rolling time the scale factor will exponentially grow up with the
Hubble parameter proportional to $\alpha^2\beta$. Thus an
exponential expansion will be insignificant unless we choose
$-\ln{h(0)}\gg 1/\alpha^2\beta$.

It is known that in inflationary models with power-law potential the
oscillating universe regime is possible when during contraction
phase the amplitude of scalar field oscillations grows up and as a
result the scalar field climbs close to the potential top. Then the
rolling down regime follows with the corresponding expansion of the
scale factor. When rolling down terminates, the oscillations starts
again and one enters a new cycle. But with the YM field present, the
probability to hit the relatively small region of the phase space
for triggering the rolling-down regime decreases, since the
cumulating chaotic oscillations and interaction with the YM
component deviate substantially the phase trajectory from
reproducing precisely the previous cycle. Therefore for small
$\alpha$ the cycling behavior demands fine tuning of the initial
data for realization of the cycling universe. Otherwise, the
evolution will terminate with a collapse to a point.

As about the YM component, its amplitude  will oscillate with a
period inversly proportional to the average value of the scalar
field <h> (i.e. of the order $\alpha^{-1}$) around the minimum
$f=-1$, since oscillations will be governed by the interaction
potential $\Vi$, while the YM potential $V_f$ will be negligibly
small. In the  case when the scalar field sits at the top of the potential
barrier (what is possible if $V_j=0$), the YM filed will oscillate
in the potential $V_f$. The oscillation period will then be
proportional to the value of the scale factor, but for sufficiently
fast increase of $a$ the frequency becomes imaginary, describing an
exponential relaxation of the YM field to the vacuum value. The scale factor
itself in this special case will be either exponentially growing, or
collapsing depending on initial conditions, This can be shown
analytically. Indeed, for $h=0$ the non-zero potential energy of the
Higgs field will play the role of the cosmological constant equal to
$\Lambda=\alpha^4\beta^2/4$. As a result, independently of the
dynamics of the YM field the equation of state will be
\begin{equation}\label{eq:sost}
    p=\frac{\eps-4\Lambda a^3}{3}.
\end{equation}
Substituting this into the continuity equation
\begin{equation}\label{eq:con}
    \dot{\eps}+3\frac{\dot{a}}{a}p=0,
\end{equation}
after trivial integration we obtain
\begin{equation}\label{eq:rhoa1}
    \eps=\Lambda a^3+\frac{C^2}{a},
\end{equation}
where $C$ - is the integration constant. For instance, for the
sphaleron configuration   $f=h=0$ with no field dynamics the energy
density is given by
\begin{equation}\label{eq:sphal2}
    \eps=V_h+V_f=\frac{\Lambda a^3}{4}+\frac{3}{2a},
\end{equation}
which results in $C^2=3/2$. Using the value (\ref{eq:rhoa1}) for the
energy density, we can solve the constraint equation for the scale
factor:
\begin{equation}\label{eq:N2}
    \dot{a}^2+1=\gamma^2 a^2/4+4\eta^2/a^2,\qd \mbox{where}\quad
    \gamma \equiv 2\alpha^2\beta\sqrt{\frac{2\pi}{3}},\quad\eta \equiv C\sqrt{\frac{2\pi}{3}}.
\end{equation}
A simple solution arises when the right hand side of this equation
is a full square. For this to happen, one has to impose the
following relation between the parameteres:
\begin{equation}\label{eq:const1}
    \gamma\eta=1.
\end{equation}
Taking the square root of the constraint equation, one obtains
\begin{equation}\label{eq:a6}
    \dot{a}=\gamma a/2-2/(\gamma a).
\end{equation}
Then the solution for the scale factor will read
\begin{equation}\label{eq:a7}
    a=\sqrt{\frac{2}{\gamma^2}+\Lef a_{0}^2-\frac{2}{\gamma^2}\Rig e^{\gamma t}},
\end{equation}
где $a(0)=a_0$. Note that the constraint equation is invariant under
the time reflection, so if $a(t)$ is a solution, then $a(-t)$ will
also be a solution. So for brevity we do not write  $\pm$ in the Eq.
(\ref{eq:a6}) as well as in the solutions for $a(t)$.

Depending on the values of the parameters, one observes either
cosmological singularity, or inflation. For $a_{0}=2/\gamma$ the
solution will be the static universe  $a(t)=a_0$: in this case the
negative pressure of the scalar field is exactly compensated by the
positive contribution of the YM field.

If the right hand side of (\ref{eq:N2}) is not a full square (which
can be expected in our Standard model scales case since the
relations  (\ref{eq:const1}) does not hold in view of  $\alpha\ll
1$), then an analytic formula for the time dependence of the scale
factor is more complicated:
\begin{equation}\label{eq:a8}
    a=\sqrt{\frac{e^{-\gamma t}}{\gamma^4\hat{C}}(1-\gamma^2\eta^2)+\frac{2}{\gamma^2}+\hat{C}e^{\gamma t}}.
\end{equation}
Here the integration constant was introduced as   $\hat{C}$ and not
as  $a(0)$. Evidently,  for small $\alpha$ the solution will be a
growing exponent with the power coefficient proportional to
$\alpha^2$. Coming back to dimensionful time parameter will get an
exponential of   $\alpha^2/T_{Pl}\sim 10^{9}\mbox{sec}^{-1}$.
Comparing this with an observed Hubble constant
$10^{-17}\mbox{sec}^{-1}$ we get  the needed value of the mass of
the scalar field $10^{-30}M_{Pl}=10^{-2}~\mbox{eV}/c^2$.

\subsection{Planck scale}
Now consider the case of  $\alpha$ and $\beta$ of the order of
unity, so all the quantities are of the Planck scale. While in the
case  $\alpha^2\beta\ll 1$ the behavior of the system was determined
by the fixed points of the Higgs potential, now we have to find the
fixed points of full system of equations   (\ref{eq:w}) and
(\ref{eq:a3}).

Equating to zero the time derivatives of the variables we get the
following system :
\begin{eqnarray}
  \frac{3h}{2a^2}(f+1)^2+h\beta^2(h^2-\alpha^2) &=& 0, \nn\\
  \frac{3h^2}{2}(f+1)+\frac{6f}{ a^2}(f^2-1) &=& 0, \nn\\
  \frac{a\beta^2}{4}(h^2-\alpha^2)^2-\frac{3}{2a^3}(f^2-1)^2 &=&
  0.\label{eq:Q0}
\end{eqnarray}
Denoting the variables collectively $Q=\{h,~f,~a\}$ we will get a
solution depending on parameters  $Q =Q_0(\alpha,\beta)$. This
solution has to satisfy the constraint equation  (\ref{eq:N}):
\begin{equation}\label{eq:kappa}
    a_0(\alpha,\beta)=\frac{8\pi}{3}\eps(\alpha,\beta).
\end{equation}
Not every solution of (\ref{eq:Q0}) does it: e.g. the vacuum state
$h=\pm \al,\: f=-1$ does not. As a result, the physical fixed points
of the system of equations are realised in the parameter space only
on the curve given by the Eq.   (\ref{eq:kappa}).

Obviously, the solution  $f=h=0$ satisfies the first two equations
of the system  (\ref{eq:Q0}), while from the third equation we find
the complete solution:
\begin{equation}\label{eq:Qsph}
    h=0,\quad f=0,\quad \al a=2/\sqrt{\btt}.
\end{equation}
He we denoted  $\btt=\beta\sqrt{8/3}$ to simplify further relations.
This static solution was already found in the previous section as
 $a_0=2/\gamma$.

Finally, the last non-trivial solution of the system (\ref{eq:Q0})
reads:
\begin{equation}\label{eq:Qw}
    \al h^2=f=\al\frac{\btt-1}{\btt+1},\quad \al a=\frac{2\sqrt{2}}{\sqrt{\btt+1}}.
\end{equation}
In the limit $\btt\ra 1+0$ this solution coinsides with the previous
one. For $\btt<1$ the solution of the type (\ref{eq:Qw}) does not
exist. One can notice that the parameter $\alpha$ us just the scale
one.

Let us try to give qualitative description of the fixed point
corresponding to the solutions described above. For the first
solution we have the following configuration.  The field $h$ sits at
the local maximum of its potential  $V_h$: $h_0=0$. This switches
off the interaction between the scalar and the YM components: $\Vi=0$.
Correspondingly, the fixed points of the YM field will be extrema of
the potential $V_f$. These are  $\pm 1,\:0$, from which only $f_0=0$
satisfies the constraint equation.

In the second case the fixed point for $h$ will be not the maximum,
but the minimum of the potential.  This is not  the minimum
$h=\pm \al$, since interaction with the YM field shifts the point of
minimum in such a way that $0<h_0<\al$. Similarly, for the YM field
the extremal points  $0,~1$ turn out to be shifted: the local
maximum to the right, and the minimum to the left according to the
relation  $f_{max/min}=(1\pm\sqrt{1-a^2h^2})/2$. From the constraint
equation we get that for  $l<3$, the fixed point is the local
maximum; for $l=3$ the maximum and the minimum coincide forming the
inflection point; with further increasing $l$ the fixed point
will be the local minimum.

Finally, the constraint equation shows in which region of the phase
space the system can be depending on parameters, or, conversely,
which should be the energy of the scalar field for the desired
regime of evolution associated with the motion near one or another
fixed points. Therefore for two families of solutions
(\ref{eq:Qsph}) and (\ref{eq:Qw}), respectively, we find the
following relations:
\begin{equation}
    \al_1=\frac{1}{\sqrt{2\pi\btt}}, \quad
    \al_2=\frac{\btt+1}{\btt\sqrt{8\pi}}.
\end{equation}

To describe evolution of the system in the vicinity of fixed points
one has to linearize the equations of motion around them. Excluding
the momentum variables, we obtain the matrix equation:
\begin{equation}
  \delta\ddot{Q}=M\delta Q,
\end{equation}
where the matrix $M$ is obtained by differentiation of the equations
of motion and the Friedmann equations over  $Q=\{h,~f,~a\}$ and
substituting the values of the variables in the fixed point. At the
same time we have to take into account the dependence between the
parameters imposed by the constraint equation. Then we obtain for
$M(l)$:
\begin{equation}\label{eq:Msph}
    M_1=\frac{1}{4\pi}\left(%
\begin{array}{ccc}
  \frac{3}{4}(\btt-1) & 0 & 0 \\
  0 & 3 & 0 \\
  0 & 0 & 1 \\
\end{array}%
\right),
\end{equation}
while in the second case:
\begin{equation}\label{eq:Mw}
    M_2= \left( \begin {array}{ccc} ,{\frac {3({\btt}^{2}-1)}{32\pi}
}& -{\frac { 3\sqrt {2} \left( \btt+1 \right)^{5/2}
}{128{\pi}^{3/2}{\btt}^{2}
  }}&{\frac {3\sqrt {2} \left( \btt+1
 \right) ^{3} \sqrt{\btt-1}}{512{
\pi}^{2}{\btt}^{2}}}
\\ \noalign{\medskip}\frac{3\sqrt{\btt-1}}{\sqrt{2\pi \left(\btt+1\right)}}&

-{\frac { 3\left( \btt-3 \right)  \left( \btt+1 \right) }{16\pi
\btt}}& -\frac{3(\btt+1)^{3/2}(\btt-1)}{32\pi^{3/2}\btt^2}
\\ \noalign{\medskip}\sqrt{\frac{\btt-1}{2}}&\frac{(\btt+1)^{3/2}(\btt-1)}
{8\sqrt{\pi}\btt^2}&{\frac { \left( \btt+1 \right) ^{2} }{16\pi{\btt}^{2}}}
\end {array} \right)
\end{equation}

It is easy to explore the case $\btt<1$, when only the matrix $M_1$
is relevant. For the scale factor the fixed point will be the
focus by $h$ and nodes by $f$ and $a$ describing an exponential
expansion or contraction of the universe. For
this we obtain the following critical value of the scale factor
$a_0=\sqrt{8\pi}$, when the inflationary potential  $V_h$ is
compensated by the deflationary potential  $V_f$ in the equation.

In order to determine the system behavior in the general case, one
has to find the eigenvalues of the linear system.
\begin{equation}\label{eq:mu}
    {\rm det}(M_{1,2}-\mu^2 \mathbb{I})=0.
\end{equation}
where the eigenvalues $\mu$ of the full system in the
six-dimensional phase space enter squared into the three-dimensional
system which is left after eliminating momenta.

The matrix $M_1$ is diagonal, so that the eigenvalues are obvious.
The eigenvalue for the scalar field is  $\mu_{h}^2=3/4(\btt-1)$. For
$\btt>1$ we have a node. For $\btt<1$, the eigenvalues become pure
imaginary, that is the node is transformed to focus.  This means that
the quadratic in $h$ interaction potential $\Vi$ dominates over the
quadratic term of the Higgs potential, so the effective potential at
the point $h=0$ has a minimum, but not a local maximum. Eigenvalues
of the gauge field and the scale factor are obviously always real.
On the phase portrait this fixed point is a node, and one therefore
has expansion or contraction of the universe. Correspondingly, the
system is unstable in the vicinity of fixed point parameterized by
the equations (\ref{eq:Qsph}).

Eigenvalue of the matrix $M_2$ can also be found analytically since
the corresponding equation (\ref{eq:mu}) is cubic in terms of
$\mu^2$. Their explicit form, however, is too long, so we do not
present it here. One root is always real, as it should be for the
cubic equation. It is negative and the corresponding fixed point is
focus. Two other roots are complex on the interval  $\btt\in
[1.8,~7.5]$. Their real parts are equal, while imaginary parts
differ in sign describing different orientation (left/right handed)
of phase trajectories. For $1<\btt<1.8$ the roots are real and
positive, while for $\btt>7.5$ --- negative. In the first case one
has nodes, in the second --- elliptic orbits.  Thus, on the interval
$\btt>7.5$ the fixed points parameterized by  (\ref{eq:Qw}), are
focal points in the full six-dimensional phase space, so the system
is stable in the vicinity of these points.

From this analysis it follows that one of the main features of the
behavior of the system on Planck scale is that apart from unstable
solution of the type of static universe, when small deviation
trigger long cycles of expansion or contraction, there are
quasi-static regimes characterized by dynamical equilibrium between
the scale factor and matter, which are stable against small
perturbations.

Finally, we discuss the role of the Higgs phase rotation parameter
$j$. When  $j\neq 0$, one observes the regimes typical for the
second family of fixed points (while the first family can not be
realized since now  $h\neq 0$). This additional parameter can be
used to set the system into the state close to stationary points.
Without it, we have some prescribed dependence between the other
parameters,  $\alpha(\beta)$, for every fixed point, which may not
be satisfied for the particular theory chosen. With $j\neq 0$,
one has more flexibility in the
parameter space to impose the desired evolution regime of the
system.

\subsection{Numerical experiments}

\begin{figure}[p]
\hbox to\linewidth{\hss%
\psfrag{x}{\LARGE{$t$}} \psfrag{f}{\LARGE{$h/\al$}}
\psfrag{h}{\LARGE{$f$}} \psfrag{a}{\LARGE{$\ln{a/a(0)}$}}
 \psfrag{w}{\LARGE{$w$}}
    \resizebox{9cm}{6cm}{\includegraphics{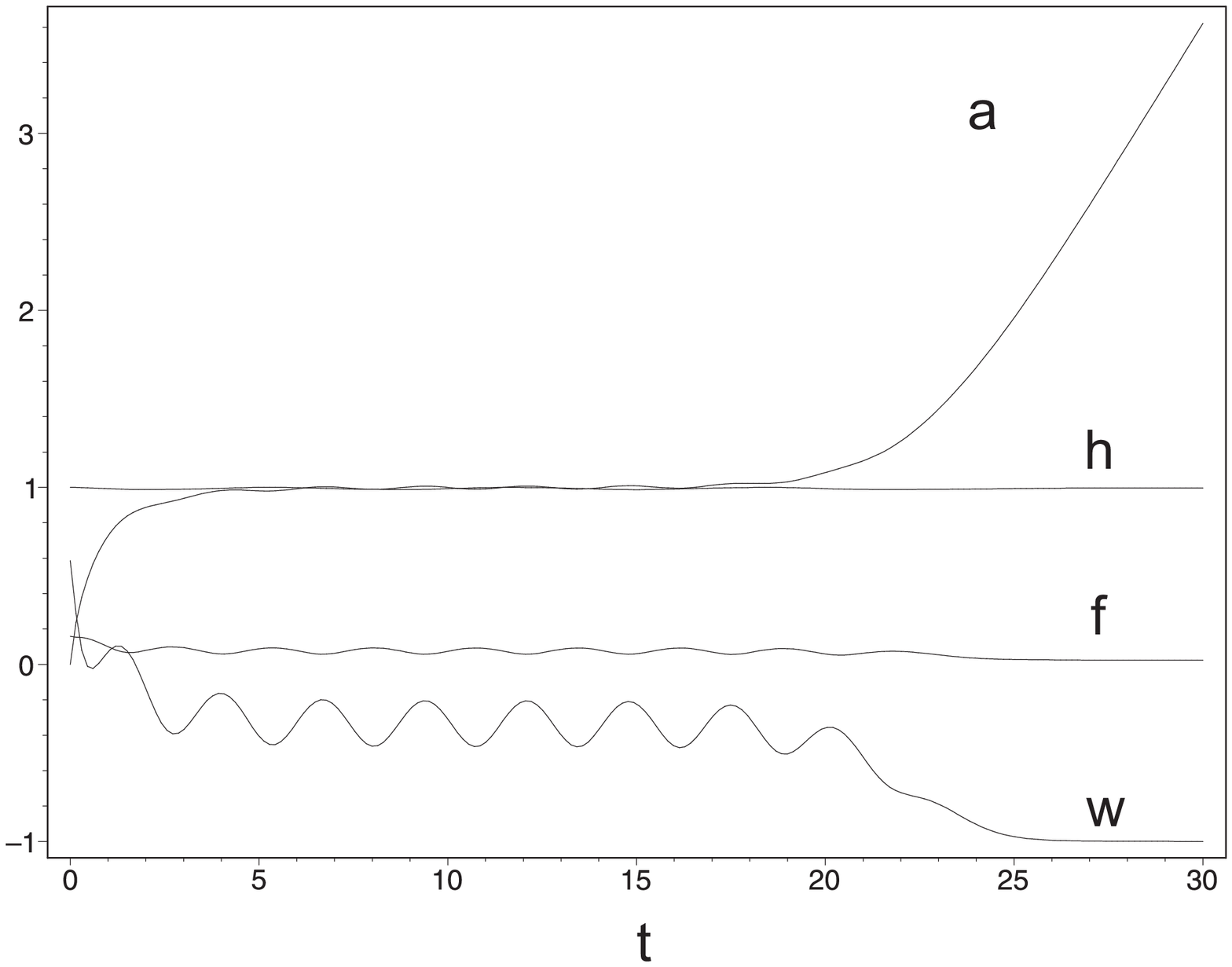}}
\hss} \caption{\small   The amplitudes $h(t)/\al,\; f(t)$ of Higgs
and YM fields and the state parameter   $w(t)=p/\eps$ for a  typical
solution with  long quasistationary state.} \label{fig:solution1}
\end{figure}

\begin{figure}[p]
\hbox to\linewidth{\hss%
\psfrag{z}{\LARGE{$\dot{a}$}}
 \psfrag{a}{\LARGE{$a$}}
  \psfrag{1}{\LARGE{$t=2$}}
 \psfrag{2}{\LARGE{$t=20$}}
    \resizebox{9cm}{6cm}{\includegraphics{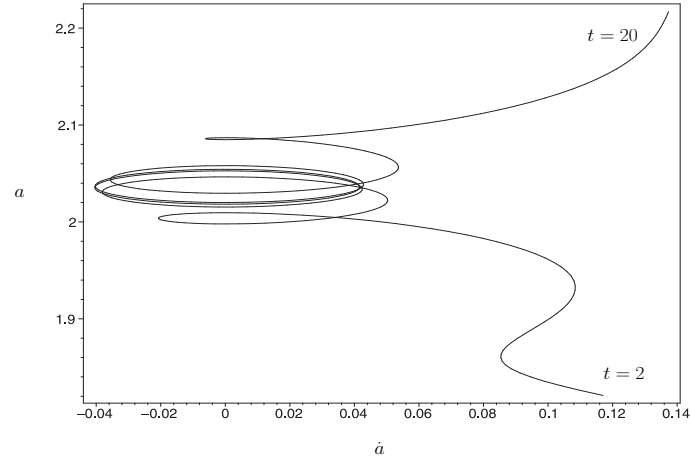}}
\hss} \caption{\small Phase trajectory for the scale factor in
vicinity of the stationary state.} \label{fig:phase_scale}
\end{figure}

\begin{figure}[p]
\hbox to\linewidth{\hss%
\psfrag{x}{\LARGE{$\dot{h}/\al$}}
 \psfrag{y}{\LARGE{$h/\al$}}
    \resizebox{9cm}{6cm}{\includegraphics{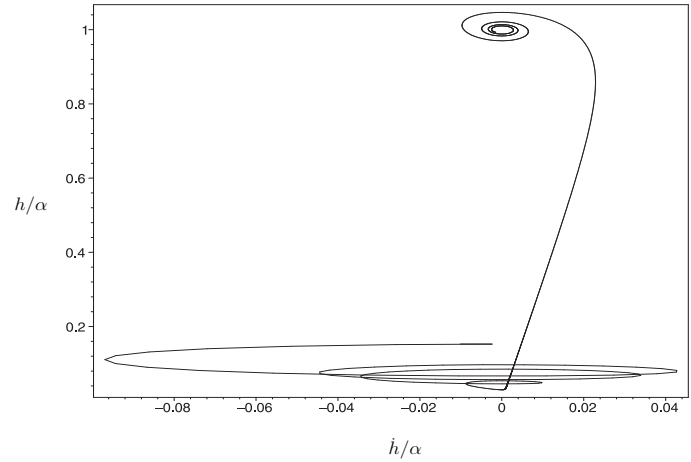}}
\hss} \caption{\small Phase tracjectory of the Higgs field
demonstrating the jump of the minimum of the scalar field potential from $h_{min}\simeq 0$ (equals to zero when $j=0$) to $\alpha$ with the
growing scale factor.} \label{fig:phase_f}
\end{figure}

Numerical solutions confirms the above qualitative considerations.
We are particularly interested in illustrating the system behavior
in different regimes described in the previous subsection. On
Figs.~(\ref{fig:solution1}--\ref{fig:phase_scale})  we present the
solution exhibiting the existence of quasi-stable state. The system
enters this state with negative and leaves it with positive cosmic
acceleration, with concrete values depending on initial conditions.
The Fig.~(\ref{fig:phase_f}) illustrates transition from the
quasi-stable local minimum near zero to a true minimum $h=\al$,
after inflationary expansion of the scale factor. These figures show
behavior of the system with Planck scale parameters, when
interaction with the YM field is sufficiently strong to change
substantially vacuum state of the scalar field.

\begin{figure}[p]
\hbox to\linewidth{\hss%
\psfrag{x}{\LARGE{$t$}} \psfrag{f}{\LARGE{$h/\al$}}
\psfrag{h}{\LARGE{$f$}} \psfrag{a}{\LARGE{$a$}}
    \resizebox{9cm}{6cm}{\includegraphics{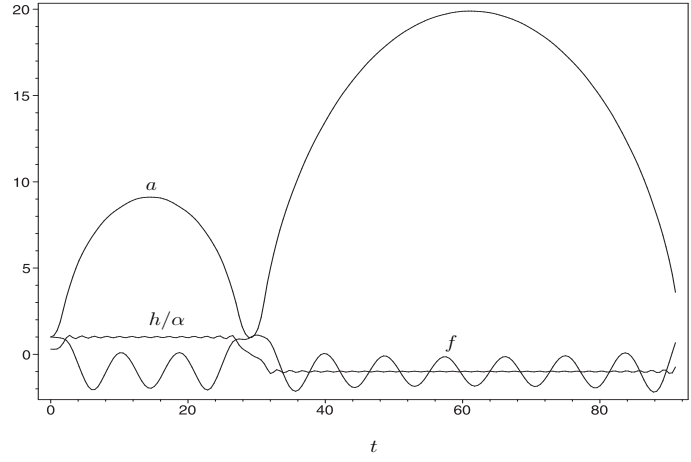}}
\hss} \caption{\small The field amplitudes and the scale factor for
two subsequent cycles with different parameters.}
\label{fig:solution_all2}
\end{figure}

The cyclic evolution is presented on Fig.~(\ref{fig:solution_all2}).
For this solution we have chosen the parameter values such that the
periods of the expansion cycles are comparable to the period of the
oscillations of the scalar field. In this case interaction between
scalar and YM fields is not strong, but still the YM field plays
crucial role at the moments of bounces. Since the oscillation
frequencies of  $h$ and $f$ are different, the vaule of the YM field
at the next bounce may differ significantly for that at the previous
bounce. With high probability the subsequent cycles differ from each
other with the YM field values being chaotically distributed at the
moments of bounces. This is the main difference of the large time
scale behavior of the EYMH system as compared with the pure scalar
cosmology.

\begin{figure}[p]
\hbox to\linewidth{\hss%
\psfrag{t}{\LARGE{$t$}} \psfrag{f}{\LARGE{$h/\al$}}
\psfrag{h}{\LARGE{$f$}}
\psfrag{a}{\LARGE{$100(\dot{a}-\dot{a}(0))$}}
    \resizebox{9cm}{6cm}{\includegraphics{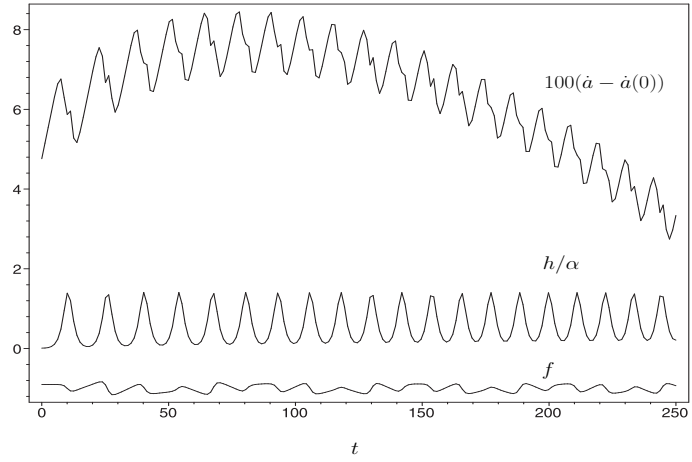}}
\hss} \caption{\small A typical solution   for small $\alpha$ (on
this plot $\alpha=10^{-6}$). Initially the scalar field amplitude is
large enough to ensure cosmic acceleration. With decreasing
amplitude an accelerated expansion  changes to contraction.  }
\label{fig:beta0}
\end{figure}

Finally, on Fig.~(\ref{fig:beta0}) we illustrate the dynamics of the
EYMH universe for asymptotically small value  $\alpha=10^{-6}$. When
the oscillation amplitude of the scalar field is relatively big (for
this figure  $h(0)=10^{-2}$), the potential term exceeds the kinetic
term leading to weakly accelerating regime. But with growing scale
factor the scalar field amplitude decreases and expansion converts
to contraction.

\subsection{EYMH hybrid inflation}
As was described in previous subsection, the EYMH system
demonstrates a wide variety of behavior in different regimes. But
the most interesting among them seems to be the next one. The EYMH
system may be considered as a very natural variant of the hybrid
inflation, introduced by Linde~\cite{Linde:1993cn}. The main idea of
the hybrid inflation was to add a massive scalar field $f$ to the
Higgs field $h$ so that the total potential reads
\begin{equation}
    V(f,h)=\frac{m^2}{2}f^2+\frac{\lambda'}{2}f^2h^2+\frac{\lambda}{4}
    (h^2-v^2)^2.
\end{equation}
In this case inflation consists of  two stages. When $f$ is larger
then the critical value $f_c=\lambda v^2/\lambda'$, the Higgs field
is trapped on the top of its potential due to the interaction
between the scalar fields. During this stage the field $f$ slowly
rolls under $f_c$ and then the second stage starts --- the roll of
$h$. Inflation ends when $h$ reaches its true minimum $|v|$.

In our system there is a similar trapping potential $\Vi$. Mention
that it does not enter the Friedmann equation for acceleration,
therefore the inflationary potential will be only $V_h$. Next, $\Vi$
depends on the gauge field and the scale factor. The gauge field
changes slowly (especially if one sets it in the minimum $f=1$ of
its potential $V_f$), but the scale factor in the denominator grows
exponentially, therefore the first stage when $h$ is trapped will be
rather short. But during the first stage the scalar field will be
lifted closer to the top of the potential $V_h$, what implies a
prolonged second stage of inflation with the slow roll of the Higgs
field. In  other words,  large inflation occurs when the initial
value of the Higgs field, $h_0$, is strictly chosen to be in
vicinity of the top: $h_0\simeq 0$. But this condition can be
weakened significantly due to the trapping potential which will
automatically prepare the Higgs field in the right position before
the rolling down.

As was mentioned above, the gauge field plays crucial role in the
trapping of the Higgs field: it switches the trapping on, when
$f_0=1$, and off, when $f_0=-1$. In the last case the Higgs field
will fall freely, reproducing the standard `slow roll' scenario. The
initial value of the scale factor  $a_0$ affects the duration of
the trapping: when $a$ reaches the critical value $a_c\simeq
\sqrt{6}/(\alpha\beta)$ (what happens rather fast due to the
exponential growth of $a$), the trapping will end.

\begin{figure}[p]
\hbox to\linewidth{\hss%
\psfrag{1}{\LARGE{$h_0/\al$}} \psfrag{2}{\LARGE{$N_{+}/60$}}
\psfrag{3}{\LARGE{$N_{-}/60$}}
\psfrag{4}{\LARGE{$N_{+}/N_{-}$}}
    \resizebox{9cm}{6cm}{\includegraphics{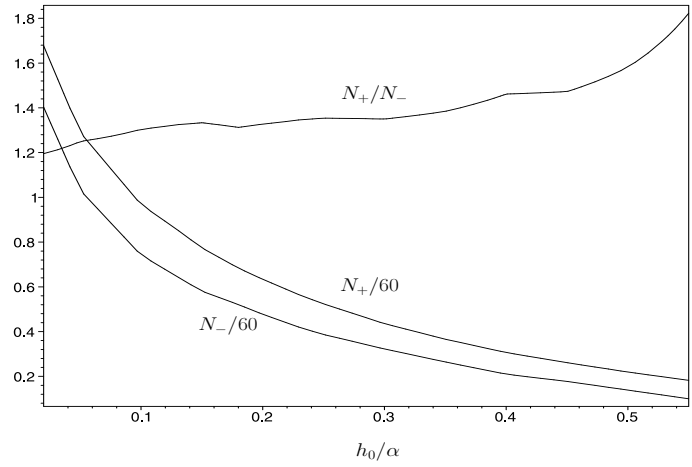}}
\hss} \caption{\small The number of $e$-folds during inflation in
the case of simple slow roll inflation, $N_{-}$, and hybrid
inflation, $N_{+}$, as functions of initial value of the scalar
field, $h_0$.} \label{fig:Nh02}
\end{figure}

Now let us turn to the numerical calculation of the number $N$ of
$e$-folds  during this hybrid inflation. On the
Fig.~(\ref{fig:Nh02}) one can see the dependence of the $N$ on the
initial value $h_0$ of the Higgs field. There are three plots:
$N_{-}$, when trapping is switched off by the gauge field $f_0=-1$;
$N_{+}$, which was calculated for the trapping with $f_0=1$ and
$a_0/a_c=1/5$; their ratio $N_{+}/N_{-}$. The valued $N_{\pm}$ are
normalized by the number $N=60$ of $e$-folds during the real
inflation of our universe \cite{Liddle:2003as}. One can see that
even for rather weak trapping with $a_0$ being just five times
smaller than the critical value, there is a $20-40\%$ gain of
$e$-folds. This gain grows with the increase of $h_0$, as expected,
because the lift of $h$ to the top of $V_h$ is significant when
$h_0$ is far from the top. Also the needed number of $e$-folds
$N\sim 60$ can be obtained in twice wider range of $h_0$ when the
trapping is on.

\begin{figure}[p]
\hbox to\linewidth{\hss%
\psfrag{1}{\LARGE{$\ln{a_0/a_c}$}} \psfrag{2}{\LARGE{$N$}}
    \resizebox{9cm}{6cm}{\includegraphics{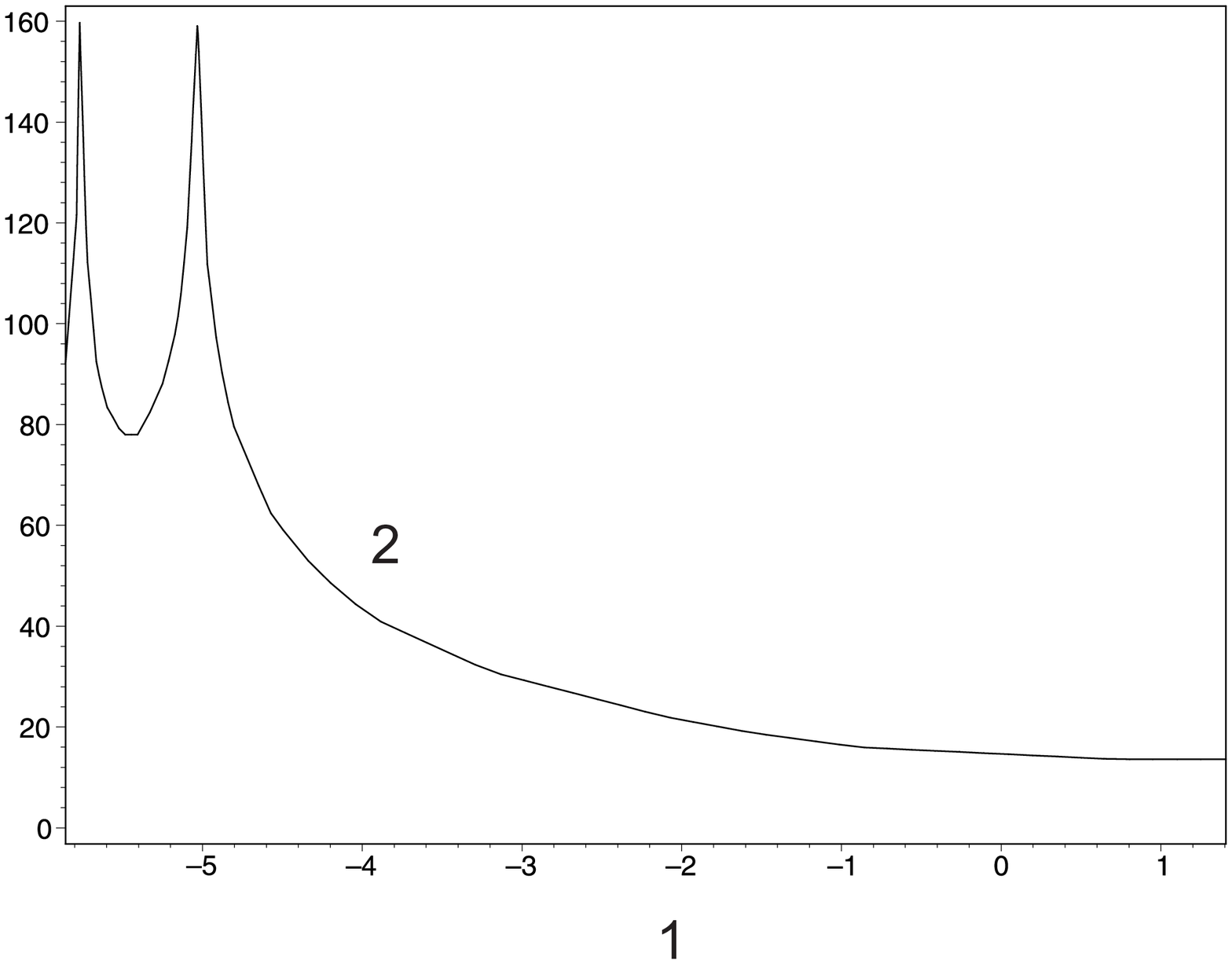}}
\hss} \caption{\small The dependence of the number $N$ of $e$-folds
on the initial value of the scale factor $a_0$, as compared with the
critical value $a_c\simeq\sqrt{6}/(\alpha\beta)$. When $a_0$ reaches
$a_c$, the hybrid inflation reduces to the simple slow roll
inflation. The peaks in the left part of the plot describe the
divergence of $N$ due to an infinite inflation, when the Higgs field
is driven exactly to the top of the Higgs potential.}
\label{fig:Na02}
\end{figure}

So switching on of the trapping due to the interaction with the
gauge field may really enhance inflation. On the
Fig.~(\ref{fig:Na02}) we can see how this amplification will depend
on the ratio $\ln{(a_0/a_c)}$. For  chosen parameters the absence of
trapping when $a_0\gg a_c$ gives us about thirteen $e$-folds. But
the choice of several $e$-folds smaller $a_0$ (so that $a_0\ll a_c$)
will allow us to gain a large total number of $e$-folds. Actually
there is even a divergence in the $e$-folds number when the trapping
potential drives the scalar field exactly on the top of the
potential $V_h$ with infinite inflation. Of cause this trapping
scenario is significant when $j=0$, and the number of $e$-folds
greatly decreases with the growth of $j$.

%\subsection{Заключение}

We conclude this section with the following remarks. Coupled
Yang-Mills-Higgs dynamics for closed FRW universe gives  rise to new
interesting evolution types which include transient regimes of
cosmic acceleration, bounces and cyclic evolution. Presence of the
YM component and Higgs phase rotation parameter changes
substantially the dynamics of the universe at small scale factors.
The interaction with the gauge field holds the Higgs field near the
top of the potential, and the balance of accelerating and
decelerating potentials of scalar and vector fields can freeze the
scale factor. The phase rotation acts in reverse, making the dynamic
of the system to be more similar to the scalar field with power-low
potential, but its kinetic nature leads to the opposite sign in the
potential at very small distances. Also it increases the number of
free parameters, which can be useful in quantitative analysis.

One intriguing feature is possibility
of an infinite sequence of cycles whose parameters change
chaotically due to evolution of the YM component. This resembles the
Multiverse models \cite{multiverse} realized as sequence of
universes with different parameters in the ultralarge time scale.

The system of interacting Higgs and YM fields can be considered as a
very natural candidate for the hybrid inflation scenario, which is
richer then a standard slow roll inflation with a scalar field. The
particular feature of the model is that due to the vector nature of
the gauge field, whose energy density depends on the scale factor,
the EYMH inflation also inherits this dependence. There can be a
large inflation in a small universe, and a small inflation in a
large universe. This looks quite similar to the current views on the
evolution of the universe with large initial inflation and slow
late-time acceleration.

\section{Non-Abelian Born-Infeld (NBI)} Open string theory suggests
the following generalization of the Maxwell Lagrangian (applicable
to any dimensions):   \be
L=\frac{\beta^2}{4\pi}\left(\sqrt{-\det(g_{\mu\nu}+
F_{\mu\nu}/\beta)}-\sqrt{-g}\right),\ee $\beta$ being the critical
BI field strength ($\beta=1/2\pi \alpha'$ in string theory). In four
dimensions this is equivalent to\be\label{sqrt}
  L=\frac{\beta^2}{4\pi}(\cR-1),\quad
 \cR=\sqrt{1+\frac{F_{\mu\nu}F^{\mu\nu}}{\beta^2}- \frac{(\tilde
F_{\mu\nu}F^{\mu\nu})^2}{16 \beta^4}}.\ee In the non-Abelian case
the strength tensor $F^{\mu\nu}$ is matrix valued, and the
prescription (Tseytlin~\cite{Ts97}) is more complicated: the
symmetrized trace, which is calculated expanding the Lagrangian in
powers of $F_a^{\mu\nu}T^a$ ($T^a$ being the gauge group
generators), then symmetrizing all products of $T^a$  involved and
only afterwards taking the trace. Symbolically this is given by the
expression \be L_{Str}=\frac{\beta^2}{4\pi} {\mathrm{Str}}
\left(\sqrt{-\det(g_{\mu\nu}+F_{\mu\nu}/\beta)}-
\sqrt{-g}\right),\ee but actually this is a useful form if one is
able to perform a subsequent resummation. Fortunately, this is
possible in the closed from for the homogeneous and isotropic SU(2)
YM field and the metric $ds^2=N^2 dt^2-a^2dl_3^2$ leading
to~\cite{Gal'tsov:2003xm}:\be L_{Str}=- N a^3
\frac{1-2K^2+2V^2-3V^2K^2}{\sqrt{1- K^2+ V^2- K^2V^2}},\quad
 K^2=\frac{\dot{w}^2}{\beta^2 a^2 N^2},\quad
V^2=\frac{(w^2-k)^2}{\beta^2 a^4}.\ee A simpler (the ordinary trace)
prescription for the NBI Lagrangian consists in summation over color
indices in the field invariants $F^a_{\mu\nu}F^{a\mu\nu},\;\; \tilde
F^a_{\mu\nu}F^{a\mu\nu}$ in the square root form of the Lagrangian
(\ref{sqrt}). This gives \be L=- N a^3 \sqrt{1-3 K^2+3 V^2- 9
K^2V^2}.\ee
\subsection{NBI cosmology}  Homogeneous and
isotropic NBI cosmology with an ordinary trace Lagrangian turns out
to be completely solvable by separation of variables~\cite{BI}.  It
leads to an interesting equation of state:\be p=\frac{\ep(\ep
_c-\ep)}{3(\ep_c+\ep)}, \ee where $ \ep_c=\beta/4\pi$  is the
critical energy density, corresponding to vanishing pressure. For
larger energies the pressure becomes negative, its limiting value
being \be p=-\ep/3.\ee This is the equation of state of an ensemble
of non-interacting isotropically distributed straight Nambu-Goto
strings (which indicates on the stringy origin of the NBI
Lagrangian). In the low-energy limit the YM equation of state $p=\ep
/3$  is recovered. Thus, the NBI FRW cosmology smoothly interpolates
between the string gas cosmology and the hot Universe. The energy
density is \be\ep=\ep_c
\;\left(\sqrt{\frac{a^4+3(w^2-k)^2}{a^4-3\dot w^4 }}-1\right).
 \ee

From the YM (NBI) equation one obtains  the following  evolution
equation
 for  the energy density:
\begin{equation}\label{drho}
     \dot {\varepsilon } =-2\frac{ \dot{a}}{a}\,\frac{\varepsilon
     \left (\varepsilon+2 \ec\right )}{\varepsilon+ \ec },
\end{equation}
which can be   integrated to give
\begin{equation}\label{aepsC}
     a^4(\varepsilon+2\ec)\varepsilon={\rm const}.
\end{equation}
 From this relation one can see that the behavior of the NBI
field interpolates between two patterns: 1)~for large energy
densities ($\varepsilon\gg \ec $) the energy density scales as
$\ve\sim a^{-2}$; 2)~for small densities $\varepsilon\ll \ec$  one
has a radiation law $\varepsilon \sim \ a^{-4}$.

Remarkably, the equation for the scale factor $a$ can be decoupled
($g=\beta G$): \be\ddot a = -\frac{2 g a (\dot a ^2+k)}{2 g a^2 + 3
(\dot a^2 + k)}\ee and admits the first integral \be 3 \left(\dot
a^2 + k \right)^2 + 4 g a^2 \left(\dot a^2 + k \right) = C,\ee which
allows to draw phase portraits for different $k$ \bei\item  {\bf
Closed.}
 The only singular point is $a=0$, $b=0,\;(b=\dot a)$ which is a
center with the eigenvalues $\pm i\sqrt{6g}$.  Solutions evolve from
left to right in the upper half-plane as time changes from~$-\infty$
to~$\infty$, and from right to left in the lower half-plane. All
solutions are of an oscillating type: they start at the singularity
($a=0$) and after a stage of expansion shrink to another
singularity. The global qualitative behavior of  solutions does not
differ substantially from that in the conformally invariant YM field
model, except near the singularity: \be
     a(t) = b_0 t - \frac{b_0 g}{9} t^3
            + O(t^5),
\ee where $b_0$ is a free parameter. Absence of the quadratic term
means that the Universe starts with zero acceleration in accord with
the equation of state  $p\approx-\ve/3$ at high densities.

\item {\bf Spatially flat.}
  There is a singular line
$b=0$ each point of which represents a solution for an empty space
(Minkowski spacetime). This set is degenerate, and there are no
solutions that reach this curve for finite values of~$a$. All
solutions in the upper half-plane after initial singularity expand
infinitely. A remarkable fact is that for this case one can write an
exact solution for $a$ in an implicit form: \be
     4 \sqrt{g} \, (t - t_0)
     = \sqrt{3} \left(\Omega - \arctan \Omega^{-1} + \pi/2 \right),
\ee where $ \Omega=\sqrt{2}\,a/\sqrt{\sqrt{a^4+C}-a^2} $. The metric
singularity is reached at $t=t_0$.

\item {\bf Open.}
  Physical domain   is $b<-1,\,b>1$. There is a center at $a=0$, $b=0$
with the eigenvalues $\pm i\sqrt{6g}$, but it lies outside the
boundary of the physical region.  Other  singular points are
$(a=0,\;\; b=\pm 1)$. These points are degenerate and cannot be
reached from any point lying in the physically allowed domain of the
phase plane. The only solutions which start from them are the
separatrices $b=\pm 1$ that represent (part of) the flat Minkowski
spacetime in special coordinates. One can easily see that all
solutions in the upper part of the physical domain $\dot a>1$ start
from the singularity and then move to
 $a \to \infty$, $ \dot a \to 1$.
\eei

\subsection{NBI on the brane}
 Replacement of the standard YM Lagrangian by the Born-Infeld one breaks
conformal symmetry, providing deviation from the hot equation of
state and creating negative pressure. Surprisingly enough, putting
the same NBI theory into the RS2 framework gives rise to an exact
restoration of the conformal symmetry by the brane non-linear
corrections~\cite{Gal'tsov:2003xm}. Choosing the ordinary trace
action  \be S=\lambda
 {\mathrm{Tr}}\int \sqrt{-\det(g_{\mu\nu}+F_{\mu\nu}/\beta)}\,d^4 x
 - \kappa^2 \int\, (R_5+2\Lambda_5)\sqrt{-g_{5}}\,d^5x,
  \ee
 where the brane tension $\lambda$ plays a role of the BI
 critical energy density, one obtains the constraint equation
  \be \left(\frac{\dot{a}}{a}\right)^2=\frac{\kappa^2}{6} \Lambda
  + \frac{\kappa^4}{36}
  (\lambda  +\ve)^2+\frac{\mathcal{E}}{a^4}-\frac{k}{a^2},\ee
  where $\mathcal{E}$ is the integration constant corresponding to the
bulk Weyl tensor projection (``dark radiation'') and, as usual, \be
\Lambda_4 = \frac12 \kappa^2(\Lambda+\frac16 \kappa^2 \lambda ^2),
  \quad  G_{(4)}=\frac{\kappa^4 \lambda  }{48 \pi }.\ee The energy
  density in this model scales as \be \ve = \lambda
  \left(\sqrt{(1+ {C}/{a^4}}-1\right),
\ee where $C$ is the integration constant. Surprisingly, the
constraint equation  comes back to that of the YM conformally
symmetric cosmology \be
  \left(\frac{\dot{a}}{a}\right)^2=\frac{8\pi G_{(4)}}{3}\Lambda_4
  +\frac{\mathcal{C}~}{a^4}-\frac{k}{a^2},\ee
where  the constant $\mathcal {C}= \mathcal{E} + \kappa^4\lambda^2
C/36$ includes contributions from both the ``dark radiation'' and
the YM energy density.

\section{Conformal symmetry breaking and dark energy}
Conformal symmetry
breaking in NBI theory demonstrates the occurrence of the negative
pressure, but its extremal value $p=-\ve/3$ is still insufficient
for DE. Meanwhile, a stronger violation of conformal symmetry may
provide an equation of state with $\ve\sim -1$. Such violation can
be of different nature: \bei \item Quantum corrections,
\item Non-minimal coupling to gravity, \item Dilaton and other
coupled scalar fields including Higgs, \item String theory
corrections.\eei Here we just explore some model Lagrangians to see
the necessary conditions for DE. Assuming the Lagrangian to be an
arbitrary function $L(\cF,\cG)$ of invariants \be
\cF=-F^a_{\mu\nu}F^{a\mu\nu}/2\;\; {\rm and}\;\; \cG=-\tilde
F^a_{\mu\nu}F^{a\mu\nu}/4,\ee one finds for the pressure and the
energy density (conformal time):

  \ba p&=&L +\left( 2\frac{\partial
L}{\partial \cF} [2(k-w^2)^2-\dot w^2]- 3\frac{\partial L}{\partial
\cG}\dot w(k-w^2)\right)a^{-4}, \non\\ \vspace{.5cm} \ve&=&-L
+\left( 6\frac{\partial L}{\partial \cF} \dot w^2+3\frac{\partial
L}{\partial \cF}\dot w(k-w^2)\right)a^{-4}.\ea
 For a simple estimate consider the power-low
dependence: \be L\sim \cF (\cF/\mu^2)^{\nu-1},\ee where $\mu$ has
the dimension of mass. Then in the electric (kinetic) dominance
regime one obtains \be W=\frac{p}{\ve}=\frac{3-2\nu}{3(2\nu-1)}.\ee
For certain $\nu$ this quantity may be arbitrarily close to $W=-1$
or even less. An electric phantom regime is thus possible.

In the magnetic (potential) dominance  regime one obtains \be
W=4\nu/3-1.\ee The value $W=-1$ can not be reached, but an
admissible DE regime is also possible. These regimes are transient
since during the evolution the electric part transforms to the
magnetic and vice versa.

Another plausible form of the lagrangian (suggested by quantum
corrections) is logarithmic~\cite{YMDE}: \be L\sim \cF
\ln(\cF/\mu^2).\ee Then the energy density is \be \ve=3\left(
T[\ln(\cF/\mu^2)+2]+V \ln(\cF/\mu^2)\right)a^{-4}, \ee and the
equation of state is \be
W=\frac{p}{\ve}=\frac{(T+V)\ln(\cF/\mu^2)+2(2V-T)}
{3(T+V)\ln(\cF/\mu^2)+6T},\ee where $T=\dot w^2,\; V=(k-w^2)^2.$ It
is easy to see that $W\sim-1$ for $\ln(\cF/\mu^2)\sim -1$. In this
case the DE regime is transient. The phantom regime is also
possible.

Thus, the DE conditions for the homogeneous and isotropic YM field
can arise indeed as a result of sufficiently strong breaking of the
conformal symmetry.

\section{Miscellanies and outlook }

Though not directly related, we would like to mention  here an
interesting attempt to derive dark energy scale cosmological
constant from gluon vacuum of QCD by Klinkhamer and Volovik
\cite{Klinkhamer:2009nn}.  The issues related to chaotic behavior of
homogeneous non-isotropic YM configurations were discussed in
\cite{chao}. Possible applications of YM fields to cosmology up to
partial identification of CMB with ``cold'' YM matter as discussed
in the Sec.~2 were recently proposed by Tipler \cite{Tipler:2007vx}.
We want, however, to conclude with more conservative viewpoint that
further work is needed to understand whether YM fields can be
relevant in realistic cosmology indeed. First of all this includes
deeper investigation of the coupled YMH dynamics and a thorough
analysis of cosmological perturbations in various EYM scenarios.

\subsection*{Acknowledgments}
We are grateful to the Organizing Committee of Slavnov Fest for
invitation, and we would like to wish Andrei Alekseevich many happy
years in physics.

 The work was supported by the RFBR  grant
08-02-01398 and Dynasty foundation.

%-----------------------------------------------------------------

\begin{thebibliography}{99}
\bibitem{Gu} A.A.Starobinsky, JETP Lett. {\bf 30}, 682 (1979); Phys. Lett. {B91},
99 (1980); A.~H.~Guth,
 ``The Inflationary Universe: A Possible Solution To The Horizon And Flatness
 Problems,''
 Phys.\ Rev.\ D {\bf 23}, 347 (1981);
 %%CITATION = PHRVA,D23,347;%%
%\bibitem{Linde:1981mu}
 A.~D.~Linde,

   ``A New Inflationary Universe Scenario: A Possible Solution Of The Horizon, Flatness,
   Homogeneity, Isotropy And Primordial Monopole Problems,''
   Phys.\ Lett.\ B {\bf 108}, 389 (1982).
   %%CITATION = PHLTA,B108,389;%%
%\bibitem{Albrecht:1982wi}
   %A.~Albrecht and P.~J.~Steinhardt,
   ``Cosmology For Grand Unified Theories With Radiatively Induced Symmetry Breaking,''
   Phys.\ Rev.\ Lett.\ {\bf 48}, 1220 (1982).
   %%CITATION = PRLTA,48,1220;%%


 %\cite{Kinney:2009vz}
\bibitem{rews}
  W.~H.~Kinney,
  ``TASI Lectures on Inflation,''
  arXiv:0902.1529 [astro-ph.CO];
%\bibitem{Baumann:2008bn}
  D.~Baumann and H.~V.~Peiris,
  ``Cosmological Inflation: Theory and Observations,''
  arXiv:0810.3022 [astro-ph];
%\bibitem{Linde:2007fr}
  A.~Linde,
  ``Inflationary Cosmology,''
  Lect.\ Notes Phys.\ {\bf 738}, 1 (2008)
  [arXiv:0705.0164 [hep-th]].
  %%CITATION = LNPHA,738,1;%%

\bibitem{DE}
%\bibitem{Copeland:2006wr}
  E.~J.~Copeland, M.~Sami and S.~Tsujikawa,
  ``Dynamics of dark energy,''
  Int.\ J.\ Mod.\ Phys.\ D {\bf 15}, 1753 (2006)
  [arXiv:hep-th/0603057];
  %%CITATION = ARXIV:0810.3022;%%
%\bibitem{Trodden:2004st}
  M.~Trodden and S.~M.~Carroll,
  ``TASI lectures: Introduction to cosmology,''
  arXiv:astro-ph/0401547.
  %%CITATION = ASTRO-PH/0401547;%%



%\cite{Ford:1989me}
\bibitem{Ford}
  L.~H.~Ford,
  ``INFLATION DRIVEN BY A VECTOR FIELD,''
  Phys.\ Rev.\ D {\bf 40}, 967 (1989).


%\cite{ArmendarizPicon:2004pm}
\bibitem{Arm}
  C.~Armendariz-Picon,
  ``Could dark energy be vector-like?,''
  JCAP {\bf 0407}, 007 (2004)
  [arXiv:astro-ph/0405267];
%\cite{Kiselev:2004py}
%\bibitem{Kiselev:2004py}
  V.~V.~Kiselev,
  ``Vector field as a quintessence partner,''
  Class.\ Quant.\ Grav.\ {\bf 21}, 3323 (2004)
  [arXiv:gr-qc/0402095];
%\cite{Wei:2006tn}
%\bibitem{Wei:2006tn}
  H.~Wei and R.~G.~Cai,
  ``Interacting vector-like dark energy, the first and second cosmological
  coincidence problems,''
  Phys.\ Rev.\ D {\bf 73}, 083002 (2006)
  [arXiv:astro-ph/0603052];
H. Wei and R. G. Cai,
``Cheng-Weyl Vector Field and its Cosmological Application,''
Journal of Cosmology and Astroparticle Physics, {\bf 0709}, 015
(2007) [astro-ph/0607064];
%\cite{Jimenez:2008au}
%\bibitem{Jimenez:2008au}
  J.~B.~Jimenez and A.~L.~Maroto,
  ``A cosmic vector for dark energy,''
  arXiv:0801.1486 [astro-ph];
%\cite{Fuzfa:2006pn}
T. S. Koivisto, and D. F. Mota,
``Vector Field Models of Inflation and Dark Energy''.
 [arXiv:0805.4229];
  J. B. Jimenez, R. Lazkoz and A. L. Maroto,
``Cosmic Vector for Dark Energy: Constraints from SN, CMB and BAO.''
[arXiv:0904.0433];

\bibitem{Linde:1979pr}
  A.~D.~Linde,
  ``Classical Yang-Mills Solutions, Condensation Of W Mesons And Symmetry Of
  Composition Of Superdense Matter,''
  Phys.\ Lett.\ B {\bf 86}, 39 (1979).


\bibitem{Galtsov:1991un}
 D.~V.~Galtsov and M.~S.~Volkov,
  ``Yang-Mills cosmology: Cold matter for a hot universe,''
  Phys.\ Lett.\ B {\bf 256}, 17 (1991).



\bibitem{YMDE}
%\bibitem{Zhao:2005bu}
  W.~Zhao and Y.~Zhang,
  ``The state equation of the Yang-Mills field dark energy models,''
  Class.\ Quant.\ Grav.\ {\bf 23}, 3405 (2006)
  [arXiv:astro-ph/0510356];
Y. Zhang, T. Y. Xia, \& W. Zhao,
``Yang-Mills Condensate Dark Energy Coupled with Matter and Radiation.
  Classical and Quantum Gravity,'' {24}, 3309 (2007) [gr-qc/0609115] ;
%\cite{Zhao:2007vn}
%\bibitem{Zhao:2007vn}
  W.~Zhao, D. Xu,
  ``Evolution of magnetic component in Yang-Mills condensate dark energy
  models,''
  Int.\ J.\ Mod.\ Phys.\ D {\bf 16}, 1735 (2007)
  [arXiv:gr-qc/0701136];
  %\cite{Bamba:2008xa}
%\bibitem{Bamba:2008xa}
  K.~Bamba, S.~Nojiri and S.~D.~Odintsov,
  ``Inflationary cosmology and the late-time accelerated expansion of the
  universe in non-minimal Yang-Mills-$F(R)$ gravity and non-minimal
  vector-$F(R)$ gravity,''
  Physical Review, {\bf D 77}, 123532 [arXiv:0803.3384];
  D.~V.~Gal'tsov,
``Non-Abelian condensates as alternative for dark energy,''
arXiv:0901.0115 [gr-qc];
 V. A. De Lorenci,
``Nonsingular and Accelerated Expanding Universe from Effective Yang-Mills Theory.''
[arXiv:0902.2672];
%\bibitem{xzplb2007}
T. Y. Xia, \& Y. Zhang,
2-loop Quantum Yang-Mills Condensate As Dark Energy,
  Physics Letters , { \bf B 656}, 19, (2007) [arXiv:0710.0077];
%\bibitem{wzxjcap2008}
S. Wang, Y. Zhang, and T. Y. Xia,
``3-loop Yang-Mills Condensate Dark Energy Model and its Cosmological Constraints.''
Journal of Cosmology and Astroparticle Physics, {\bf 10}, 037(2008)
[arXiv:0803.2760];
%\bibitem{zxijmpd2007}
W. Zhao, \& D. H. Xu,
``Evolution of Magnetic Component in Yang-Mills Condensate Dark Energy Models.
 International Journal of Modern Physics,'' {\bf D 16}, 1735 (2007) [gr-qc/0701136];
%\bibitem{zijmpd2008}
W. Zhao,
``Statefinder Diagnostic for Yang-Mills Dark Energy Model.
 International Journal of Modern Physics,'' {\bf D 17}, 1245 (2008) [arXiv:0711.2319];
%\bibitem{tzxijmpd2009}
M. L. Tong, Y. Zhang, \& T. Y. Xia,
``Statefinder Parameters for Quantum Effective Yang-Mills Condensate Dark Energy Model.
 International Journal of Modern Physics,'' {\bf D 18}, 797 (2009) [arXiv:0809.2123];
W. Zhao,
``Attractor Solution in Coupled Yang-Mills Field Dark Energy Models.
 International Journal of Modern Physics,''
{\bf D18}, 1331 (2009) [arXiv:0810.5506];
%\cite{Zhao:2009wy}
%\bibitem{Zhao:2009wy}
  W.~Zhao, Y.~Zhang and M.~L.~Tong,
  ``Quantum Yang-Mills Condensate Dark Energy Models,''
  arXiv:0909.3874 [astro-ph.CO].
  %%CITATION = ARXIV:0909.3874;%%

\bibitem{Cervero:1978db}
  J.~Cervero and L.~Jacobs,
  ``Classical Yang-Mills Fields In A Robertson-Walker Universe,''
  Phys.\ Lett.\ B {\bf 78}, 427 (1978).
%\cite{Henneaux:1982vs}
 \bibitem{Henneaux:1982vs}
  M.~Henneaux,
  ``Remarks On Space-Time Symmetries And Nonabelian Gauge Fields,''
  J.\ Math.\ Phys.\ {\bf 23}, 830 (1982);
  %\cite{Henneaux:1982vs}
\bibitem{Hosotani:1984wj}
  Y.~Hosotani,
  ``Exact Solution To The Einstein Yang-Mills Equation,''
  Phys.\ Lett.\ B {\bf 147}, 44 (1984);

%\cite{Moniz:1990hf}
\bibitem{Moni}
  P.~V.~Moniz and J.~M.~Mourao,
 ``Homogeneous and isotropic closed cosmologies with a gauge sector,''
  Class.\ Quant.\ Grav.\ {\bf 8}, 1815 (1991);
%\bibitem{Bertolami:1991ff}
  O.~Bertolami, Yu.~A.~Kubyshin and J.~M.~Mourao,
  ``Stability of compactification in Einstein Yang-Mills theories after
  %inflation,''
  Phys.\ Rev.\ D {\bf 45}, 3405 (1992);
%\cite{Moniz:1991kx}
%\bibitem{Moniz:1991kx}
  P.~V.~Moniz, J.~M.~Mourao and P.~M.~Sa,
  ``The Dynamics Of A Flat Friedmann-Robertson-Walker Inflationary Model In The
  Presence Of Gauge Fields,''
  Class.\ Quant.\ Grav.\ {\bf 10}, 517 (1993);
%\cite{Cavaglia:1993en}
%\bibitem{Cavaglia:1993en}
  M.~Cavaglia and V.~de Alfaro,
  ``On a quantum universe filled with Yang-Mills radiation,''
  Mod.\ Phys.\ Lett.\ A {\bf 9}, 569 (1994)
  [arXiv:gr-qc/9310001]; %\cite{Bertolami:1994jn}
%\bibitem{Bertolami:1994jn}
  O.~Bertolami and P.~V.~Moniz,
  ``Decoherence of Friedmann-Robertson-Walker geometries in the presence of
  massive vector fields with U(1) or SO(3) global symmetries,''
  Nucl.\ Phys.\ B {\bf 439}, 259 (1995)
  [arXiv:gr-qc/9410027];
%\cite{Kapetanakis:1994ip}
%\bibitem{Kapetanakis:1994ip}
  D.~Kapetanakis, G.~Koutsoumbas, A.~Lukas and P.~Mayr,
  ``Quantum cosmology with Yang-Mills fields,''
  Nucl.\ Phys.\ B {\bf 433}, 435 (1995)
  [arXiv:hep-th/9403131];
  %%CITATION = NUPHA,B433,435;%%
%\cite{Bento:1994dw}
%\bibitem{Bento:1994dw}
  M.~C.~Bento and O.~Bertolami,
  ``General Cosmological Features Of The Einstein Yang-Mills Dilaton System In
  String Theories,''
  Phys.\ Lett.\ B {\bf 336}, 6 (1994)
  [arXiv:gr-qc/9405038]; %\cite{Cavaglia:1994zv}
%\bibitem{Cavaglia:1994zv}
  M.~Cavaglia, V.~De Alfaro and A.~T.~Filippov,
  ``Quantization of the Robertson-Walker Universe,''
%\href{http://www.slac.stanford.edu/spires/find/hep/www?irn=7188447}{SPIRES entry}
{\it in Proc. Quantum Systems: New Trends And Methods} (QS 94) 23-29
May 1994, Minsk, Belarus, pp. 31-46
  J.\ Math.\ Phys.\ {\bf 38}, 4696 (1997)
  [arXiv:gr-qc/9610026];
%\cite{Moniz:1996ja}
%\bibitem{Moniz:1996ja}
  P.~V.~Moniz,
  ``Quantization of a Friedmann-Robertson-Walker Model with Gauge Fields in N=1
  Supergravity,''
  arXiv:gr-qc/9604045;
%\cite{Moniz:1996uc}
%\bibitem{Moniz:1996uc}
 % P.~V.~Moniz,
  ``FRW model with vector fields in N=1 supergravity,''
  Helv.\ Phys.\ Acta {\bf 69}, 293 (1996).

%\cite{Kuenzle:1991wa}
\bibitem{Ku}
  H.~P.~Kuenzle,
  ``SU(n) Einstein Yang-Mills fields with spherical symmetry,''
  Class.\ Quant.\ Grav.\ {\bf 8}, 2283 (1991);\\
  %\cite{Darian:1996mb}
%\bibitem{Darian:1996mb}
  B.~K.~Darian and H.~P.~Kunzle,
  ``Cosmological Einstein-Yang-Mills equations,''%\cite{Bertolami:1990iw}
%

\bibitem{Inst}
  A.~Fuzfa,
  ``Gravitational instability of Yang-Mills cosmologies,''
  Class.\ Quant.\ Grav.\ {\bf 20}, 4753 (2003)
  [arXiv:gr-qc/0310032];
%\bibitem{zraa2009}
W. Zhao,
``Perturbations of the Yang-Mills Field in the Universe,''
Research in Astronomy and Astrophysics , {\bf 9}, 874 (2009)
[astro-ph/0508010]; J. B. Jimenez, T. S. Koivisto, A. L. Maroto and
D. F. Mota,
``Perturbations in electromagnetic dark energy.''
[arXiv:0907.3648].


\bibitem{Gist}
  G.~W.~Gibbons and A.~R.~Steif,
 ``Yang-Mills cosmologies and collapsing gravitational sphalerons,''
  Phys.\ Lett.\ B {\bf 320}, 245 (1994)
  [arXiv:hep-th/9311098];
%\bibitem{Volkov:1993gp}
  M.~S.~Volkov,
  ``Einstein Yang-Mills sphalerons and fermion number nonconservation,''
  Phys.\ Lett.\ B {\bf 328}, 89 (1994)
  [arXiv:hep-th/9312005];
 %\bibitem{Volkov:1996hm}
  M.~S.~Volkov,
  ``Computation of the winding number diffusion rate due to the cosmological
  sphaleron,''
  Phys.\ Rev.\ D {\bf 54}, 5014 (1996)
  [arXiv:hep-th/9604054] ;
 %\cite{Ding:1994nw}
%\bibitem{Ding:1994nw}
  S.~X.~Ding,
  ``Cosmological sphaleron from real tunneling and its fate,''
  Phys.\ Rev.\ D {\bf 50}, 3755 (1994)
  [arXiv:gr-qc/9407036].
  %%CITATION = PHRVA,D50,3755;%%
%\cite{Fuzfa:2003gc}

  %\cite{Volkov:1998cc}
\bibitem{VoGa}
  M.~S.~Volkov and D.~V.~Gal'tsov,
  ``Gravitating non-Abelian solitons and black holes with Yang-Mills fields,''
  Phys.\ Rept.\ {\bf 319}, 1 (1999)
  [arXiv:hep-th/9810070].

\bibitem{worm}
%\bibitem{Hosoya:1989zn}
  A.~Hosoya and W.~Ogura,
  ``Wormhole Instanton Solution In The Einstein Yang-Mills System,''
  Phys.\ Lett.\ B {\bf 225}, 117 (1989);
%\cite{Das:1989ne}
%\bibitem{Das:1989ne}
  A.~K.~Das and J.~Maharana,
  ``Wormhole solution in coupled Yang-Mills axion system,''
  Phys.\ Rev.\ D {\bf 41}, 699 (1990);
%\cite{Rey:1989th}
%\bibitem{Rey:1989th}
  S.~J.~Rey,
  ``Space-time wirmholes with Yang-Mills fields,''
  Nucl.\ Phys.\ B {\bf 336}, 146 (1990);
 %\cite{Gupta:1989bs}
%\bibitem{Gupta:1989bs}
  A.~K.~Gupta, J.~Hughes, J.~Preskill and M.~B.~Wise,
  ``Magnetic Wormholes And Topological Symmetry,''
  Nucl.\ Phys.\ B {\bf 333}, 195 (1990);
%\bibitem{Berto-eu}
  O.~Bertolami and J.~M.~Mourao,
  ``Euclideanized Einstein Yang-Mills Equations, Wormholes And The Ground State
  Wave Function Of A Radiation Dominated Universe,''
%\href{http://www.slac.stanford.edu/spires/find/hep/www?irn=2594137}{SPIRES entry}
{\it In Lisbon 1990, Proceedings, The physical universe* 21-38
(QB981:A9:1990)};
%\cite{Berto}
%\bibitem{Bertolami:1990je}
  O.~Bertolami, J.~M.~Mourao, R.~F.~Picken and I.~P.~Volobuev,
  ``Dynamics of euclidenized Einstein Yang-Mills systems with arbitrary gauge
  groups,''
  Int.\ J.\ Mod.\ Phys.\ A {\bf 6}, 4149 (1991);
%\cite{Verbin:1989sg}
\bibitem{Verb}
  Y.~Verbin and A.~Davidson,
 ``Quantized Nonabelian Wormholes,''
  Phys.\ Lett.\ B {\bf 229}, 364 (1989);

  \bibitem{Don}
E.~E.~Donets and D.~V.~Galtsov,
 ``Continuous family of Einstein Yang-Mills wormholes,''
  Phys.\ Lett.\ B {\bf 294}, 44 (1992);
  [arXiv:gr-qc/9209008];\
 %\cite{Donets:1992vx}
%\bibitem{Donets:1992vx}
  E.~E.~Donets and D.~V.~Galtsov,
  ``Wormhole solutions in coupled Einstein Yang-Mills axion system,''
%\href{http://www.slac.stanford.edu/spires/find/hep/www?irn=3236145}{SPIRES entry}
{\it In *Evora 1992, Proceedings, Classical and quantum gravity*
289-292};
%\cite{Lukas:1994qf}
%\bibitem{Lukas:1994qf}
  A.~Lukas,
  ``Wormhole effects on Yang-Mills theory,''
  Nucl.\ Phys.\ B {\bf 442}, 533 (1995)
  [arXiv:gr-qc/9407037].
  %%CITATION = NUPHA,B442,533;%%

\bibitem{worm2}
%\cite{Kim:2000mu}
%\bibitem{Kim:2000mu}
  H.~Kim and Y.~Yoon,
  ``Effects of gravitational instantons on Yang-Mills instanton,''
  Phys.\ Lett.\ B {\bf 495}, 169 (2000)
  [arXiv:hep-th/0002151];
  %%CITATION = PHLTA,B495,169;%%
%\cite{Kim:2000mg}
%\bibitem{Kim:2000mg}
  H.~Kim and Y.~Yoon,
  ``Instanton-meron hybrid in the background of gravitational instantons,''
  Phys.\ Rev.\ D {\bf 63}, 125002 (2001)
  [arXiv:hep-th/0012055]; %\cite{VargasMoniz:2002gj}
%\bibitem{VargasMoniz:2002gj}
  P.~Vargas Moniz,
  ``FRW wormhole instantons in the non-Abelian Born-Infeld theory,''
  Phys.\ Rev.\ D {\bf 66}, 064012 (2002) ;
  %%CITATION = PHRVA,D66,064012;%%
%%CITATION = PHRVA,D63,125002;%%
%\cite{Mosna:2009gc}
%\bibitem{Mosna:2009gc}
  R.~A.~Mosna and G.~M.~Tavares,
  ``New self-dual solutions of SU(2) Yang-Mills theory in Euclidean
  Schwarzschild space,''
  Phys.\ Rev.\ D {\bf 80}, 105006 (2009)
  [arXiv:0909.5145 [math-ph]];
  %%CITATION = PHRVA,D80,105006;%%
%\cite{Cherkis:2009jm}

%\cite{Breitenlohner:2000rp}
\bibitem{bfm}
  P.~Breitenlohner, P.~Forgacs and D.~Maison,
  ``Static cosmological solutions of the Einstein-Yang-Mills-Higgs
  equations,''
  Phys.\ Lett.\  B {\bf 489}, 397 (2000)
  [arXiv:gr-qc/0006046].
  %%CITATION = PHLTA,B489,397;%%

%\cite{Linde:1993cn}
\bibitem{Linde:1993cn}
  A.~D.~Linde,
  ``Hybrid inflation,''
  Phys.\ Rev.\  D {\bf 49} (1994) 748
  [arXiv:astro-ph/9307002].
  %%CITATION = PHRVA,D49,748;%%


%\cite{Liddle:2003as}
\bibitem{Liddle:2003as}
%\cite{Liddle:2000cg}
%\bibitem{Liddle:2000cg}
  A.~R.~Liddle and D.~H.~Lyth,
  ``Cosmological inflation and large-scale structure,'' Cambridge University
Press (2000).
  %%CITATION = ISBN-13-9780521828499;%%
 A.~R.~Liddle and S.~M.~Leach,
  ``How long before the end of inflation were observable perturbations
  produced?'',
  Phys.\ Rev.\  D {\bf 68}, 103503 (2003)
  [arXiv:astro-ph/0305263].
  %%CITATION = PHRVA,D68,103503;%%


%\cite{Tegmark:2003sr}
\bibitem{multiverse}
%\bibitem{Tegmark:2003sr}
  M.~Tegmark,
  ``Parallel universes,''
  [arXiv:astro-ph/0302131].
  %%CITATION = ASTRO-PH/0302131;%%

%\cite{Tegmark:2003db}
%\bibitem{Tegmark:2003db}
  M.~Tegmark,
  ``Parallel universes. Not just a staple of science fiction, other universes
  are a direct implication of cosmological observations,''
  Sci.\ Am.\  {\bf 288N5} (2003) 30
  [Spektrum Wiss.\  {\bf 2003N8} (2003) 34].
  %%CITATION = 00181,2003N8,34;%%

%\cite{Weinberg:2005fh}
%\bibitem{Weinberg:2005fh}
  S.~Weinberg,
  ``Living in the multiverse,''
  arXiv:hep-th/0511037.
  %%CITATION = HEP-TH/0511037;%%

\bibitem{Ts97}
A.~A. Tseytlin. ``On non-abelian generalisation of the
{B}orn-{I}nfeld action in string
  theory''
 { Nucl. Phys.}, {\bf B501}, 41--52 (1997).

%\cite{Dyadichev:2001su}
\bibitem{BI}
  V.~V.~Dyadichev, D.~V.~Gal'tsov, A.~G.~Zorin and M.~Y.~Zotov,
  ``Non-Abelian Born-Infeld cosmology,''
  Phys.\ Rev.\ D {\bf 65}, 084007 (2002)
  [arXiv:hep-th/0111099];
\bibitem{Gal'tsov:2003xm}
  D.~V.~Gal'tsov and V.~V.~Dyadichev,
  ``Non-Abelian brane cosmology,''
  Astrophys.\ Space Sci.\ {\bf 283}, 667 (2003);
  [arXiv:hep-th/0301044];
%\cite{Fuzfa:2005qn}
\bibitem{Fuzfa:2005qn}
  A.~Fuzfa and J.~M.~Alimi,
 ``Non-Abelian Einstein-Born-Infeld dilaton cosmology,''
  Phys.\ Rev.\ D {\bf 73}, 023520 (2006)
  [arXiv:gr-qc/0511090];
  A.~Fuzfa and J.~M.~Alimi,
  ``Dark energy as a Born-Infeld gauge interaction violating the equivalence
  principle,''
  Phys.\ Rev.\ Lett.\ {\bf 97}, 061301 (2006)
  [arXiv:astro-ph/0604517];
%\cite{Novello:2006ng}
\bibitem{Novello:2006ng}
  M.~Novello, E.~Goulart, J.~M.~Salim and S.~E.~Perez Bergliaffa,
  ``Cosmological effects of nonlinear electrodynamics,''
  Class.\ Quant.\ Grav.\ {\bf 24}, 3021 (2007)
  [arXiv:gr-qc/0610043].
E. Elizalde, J. E. Lidsey, S. Nojiri, \& S. D. Odintsov, (2003).
Born-Infeld Quantum Condensate as Dark Energy in the Universe. {\it
Physics Letters B}, 571, 1; [hep-th/0307177]

%\cite{Klinkhamer:2009nn}
\bibitem{Klinkhamer:2009nn}
  F.~R.~Klinkhamer and G.~E.~Volovik,
  ``Gluonic vacuum, q-theory, and the cosmological constant,''
  Phys.\ Rev.\  D {\bf 79}, 063527 (2009)
  [arXiv:0811.4347 [gr-qc]].
  %%CITATION = PHRVA,D79,063527;%%
%\cite{Tipler:2007vx}

\bibitem{chao}
%\bibitem{Gal'tsov:2003gx}
  D.~V.~Gal'tsov and V.~V.~Dyadichev,
  ``Stabilization of the Yang-Mills chaos in non-Abelian Born-Infeld theory,''
  JETP Lett.\ {\bf 77}, 154 (2003)
  [Pisma Zh.\ Eksp.\ Teor.\ Fiz.\ {\bf 77}, 184 (2003)]
  [arXiv:hep-th/0301069];
%\bibitem{Dyadichev:2004ix}
  V.~V.~Dyadichev, D.~V.~Gal'tsov and P.~Vargas Moniz,
 ``Chaos - order transition in Bianchi I non-Abelian Born-Infeld cosmology,''
  Phys.\ Rev.\ D {\bf 72}, 084021 (2005)
  [arXiv:hep-th/0412334].
%\cite{Dyadichev:2006xc}
%\bibitem{Dyadichev:2006xc}
  V.~V.~Dyadichev, D.~V.~Galtsov and P.~V.~Moniz,
  ``New features about chaos in Bianchi I non-Abelian Born-Infeld cosmology,''
  AIP Conf.\ Proc.\ {\bf 861}, 312 (2006).

\bibitem{Tipler:2007vx}
  F.~J.~Tipler,
  ``Feynman-Weinberg Quantum Gravity and the Extended Standard Model as a
  Theory of Everything,''
  Rept.\ Prog.\ Phys.\  {\bf 68}, 897 (2005)
  [arXiv:0704.3276 [hep-th]].
  %%CITATION = RPPHA,68,897;%%


\end{thebibliography}
\end{document}